\documentclass[12pt]{iopart}
\usepackage{graphicx}
\begin{document}
\title[Quantum Correlations]{Quantum Correlations Between Identical and Unidentical Atoms in a Dissipative Environment}
\author{Ferdi Altintas and Resul Eryigit}
\address{Department of Physics, Abant Izzet Baysal University, Bolu, 14280-Turkey.}
\ead{resul@ibu.edu.tr}
\begin{abstract}
We have studied the dynamics of quantum correlations such as entanglement, Bell-nonlocality and quantum discord between identical as well as unidentical atoms interacting with a single-mode cavity field and subject to  cavity decay. The effect of single atom detuning, cavity decay rate and initial preparation of the atoms on the corresponding correlation measures have been investigated. It is found that even under strong dissipation, time evolution can create high quantum discord while entanglement and Bell nonlocality stay zero for an initially separable state. Quantum discord increases while entanglement decreases in a certain time period under dissipation for the initial state that both atoms are in the excited state if the qubits are identical. For some type of initial states, cavity decay is shown to drive the system to a stationary state with high entanglement and quantum discord. 
\end{abstract}
\pacs{03.65.Ud, 03.65.Yz, 03.67.-a}

\maketitle
\section{Introduction}
Entanglement, nonlocality and quantum discord~(QD) are all different but somehow related aspects of the quantumness of the correlated composite quantum systems. Entanglement is a kind of quantum correlation and determines whether the given state is separable or not~\cite{ld}. On the other hand, the quantum systems may contain quantum correlations other than entanglement. Quantum discord, which is defined as the difference between the quantum versions of  two analogous classically equivalent definitions of mutual information, captures all types of nonclassical correlations including entanglement~\cite{howhz}. Although entanglement and QD are equivalent for pure states~\cite{ara}, QD is found to be nonzero for some separable states~\cite{ara,lhvv}. Such states were shown to be useful in the deterministic quantum computations with one pure qubit context~\cite{dsc,ekrl}. Therefore QD is believed a new resource for quantum computation. The nonlocality as measured by the violation of Bell's inequalities is also a signature of the inseparability of a quantum state and signifies entangled states whose correlations cannot be reproduced by a classical local hidden-variable model~\cite{bell}. However it must be emphasized here that the presence of entanglement does not always imply violation of Bell's inequalities, while the violation of Bell inequalities implies entanglement. Such entangled states violating Bell's inequalities play a central role in some applications in quantum information science~\cite{agms}, such as to guarantee the safety of device-independent key distrubition protocols in quantum cryptography~\cite{agm,ngrt}.

The relations among different measures of quantum correlations and their dynamical behaviour have been an active area of research with an aim to understand the fundamental question of what quantumness is as well as to characterize these correlations as useful resources for various quantum tasks~\cite{fare1,sls,lxsw,wsfb,fwbac,wxcf,wtrr,gglxg,twgr,pljs,mpm,fpm,hms,wlnl,fcc,lyf,zcac,ykl,kk,ll,fare2,mbfc,ldlj,bfc,bfc2,werner}. One of the most important obstacle to realizing quantum operations is the interaction between the qubits and the environment which tends to wipe out the quantumness of the system~\cite{ld}. Characterization of the effect of environment on the system such as decoherence and dissipation is important. The dynamics of quantum correlations such as entanglement, quantum discord and Bell nonlocality for a two-qubit system subject to various environments have been studied by many groups recently~\cite{fare1,sls,lxsw,wsfb,fwbac,wxcf,wtrr,gglxg,twgr,pljs,mpm,fpm,hms,wlnl,fcc,lyf,zcac,ykl,kk,ll,fare2,mbfc,ldlj,bfc,bfc2,werner}. Entanglement sudden death~(ESD) is one of the new concepts that arise within the context of these studies~\cite{tyjhe}. ESD refers to death of entanglement between two qubits in a finite time while the coherence of the single qubit decays only exponentially. In contrary to entanglement, QD is more robust in a decohering environment, actually it was shown that almost all states have non-zero quantum discord~\cite{faccc}. QD was shown to decay exponentially in Markovian environments, even at finite temperatures~\cite{wsfb,wxcf} and can become zero momentarily under non-Markovian dynamics~\cite{fwbac}. QD was also experimentally investigated in an all-optical set up recently~\cite{xxcf}. Furthermore, QD and entanglement were analyzed for a Heisenberg spin chain with quantum phase transitions and QD was found to detect the critical points of the transition at finite temperatures while entanglement could  not~\cite{wtrr}. Very recently, Sun {\it et al.} investigated the induced quantum correlations between two qubits coupled to a common environment modelled by Ising spin chain in a transverse field and during the time evolution, non-zero quantum discord is found to be created for specific initial states whereas entanglement cannot be induced~\cite{sls}. In Ref.~\cite{mpm} quantum discord is found to be totally unaffected by non-dissipative noise for a long time interval under certain conditions, while entanglement decays to zero quickly. These remarkable properties of quantum discord makes it more practical than the entanglement to characterize quantum correlations.  

As mentioned, the nonlocality is curicial for some applications in quantum information. Therefore, many theoretical studies have been devoted to the comparision of entanglement and the violations of Bell inequalities~\cite{kk,ll,fare2,mbfc,ldlj,bfc,bfc2,werner}. It was  demonstrated by several authors that the survival time for entanglement is much larger than that of the Bell-inequality violation~\cite{kk,ll,fare2,bfc2}. In Ref.~\cite{bfc}, the nonlocal entanglement identified by the violations of a Bell inequality is found to be preserved during the time evolution of a system consisting of two qubits which are embedded independently in zero-temperature bosonic reservoirs and are initially in an entangled mixed state. Bellomo {\it et al.} showed that Bell inequality might not be violated for a state with high entanglement for a two-qubit system subject to amplitude damping~\cite{bfc2}. Moreover, the nonlocal entanglement between two qubits subject to classical dephasing environment modelled as Ornstein-Uhlenbeck noise has been investigated in Refs.~\cite{ll,fare2}, very recently, and it was demonstrated that the strong non-Markovian effect can protect the nonlocal entanglement~\cite{ll,fare2}. Additionally, the nonlocal entanglement induced by the qubit-qubit interaction is found to be protected against the sudden death despite the strong influence of the environmental noise for a certain class of initial states regardless of whether the environment is Markovian or non-Markovian~\cite{fare2}.  

Although the dynamical relation between the entanglement and quantum discord as well as entanglement and Bell nonlocality for a bipartite system subject to different environments have been investigated by a large number of groups as mentioned above, further investigations would help in clarifying the relation of these three correlation measures.  Along these lines, in the present paper, we have investigated the dynamics of quantum correlations such as quantum discord, entanglement and Bell nonlocality between two qubits in a lossy cavity. Very recently, entanglement, Bell nonlocality and quantum discord have been studied in cavity QED~\cite{fwbac,wxcf,gglxg,pljs,fpm,wlnl,lyf,mbfc,bfc,bfc2,mmsg}. Most of them assume identical atoms, or the atom-cavity resonant case, or only one excitation. Here, we have investigated the effect of atom-cavity detuning and cavity decay rate on the corresponding correlation measures  for a system which is initially prepared in  separable or  maximally entangled state. The most important results of the present study are: depending on the initial preparation of the system and the type of the detunings between the frequency of the cavity mode and the transition frequency of the qubits, the induced quantum discord is found to increase for a long time under cavity decay loss mechanism for certain conditions, while there is no increase in entanglement and Bell nonlocality. In some situations, the quantum discord is induced even while the atoms remain separable at all times. Also, the preservation of induced entanglement and quantum discord was observed under strong dissipation for certain type of initial states.

The paper  is organized as follows. In Sec.~\ref{model}, the model is introduced and the exact differential equations governing the dynamics of two atoms and single cavity mode are derived. The correlation measures; concurrence, CHSH-Bell inequality and quantum discord are also introduced, briefly. In Secs.~\ref{cavity1} and~\ref{cavity2}, the effect of cavity damping on the corresponding correlation measures are revealed for different detunings  for the atoms initially prepared in  separable or  maximally entangled state. The summary of the important results is given in Sec.~\ref{conclusion}.

                                                                                                                                              \section{The Model and Correlation Measures}\label{model}
Here we consider two two-level atoms with excited states $\left|e_A\right\rangle, \left|e_B\right\rangle$ and ground states $\left|g_A\right\rangle, \left|g_B\right\rangle$ interacting with a single-mode cavity field of frequency $\omega$. The Hamiltonian of this system is given as~\cite{pkpgsa}:
\begin{eqnarray}\label{hamiltonian}
H&=&\frac{\hbar\omega_A}{2}(\left|e_A\right\rangle\left\langle e_A\right|-\left|g_A\right\rangle\left\langle g_A\right|)+\frac{\hbar\omega_B}{2}(\left|e_B\right\rangle\left\langle e_B\right|-\left|g_B\right\rangle\left\langle g_B\right|)+\omega a^{\dagger}a\nonumber\\
&+&\hbar g_1(a\left|e_A\right\rangle\left\langle g_A\right|+a^{\dagger}\left|g_A\right\rangle\left\langle e_A\right|)+\hbar g_2(a\left|e_B\right\rangle\left\langle g_B\right|+a^{\dagger}\left|g_B\right\rangle\left\langle e_B\right|),
\end{eqnarray}
where $\omega_A$ and $\omega_B$ are the transition frequencies for the atoms $A$ and $B$, $a$ and $a^{\dagger}$ are the annihilation and creation operators for the field and $g_1$ and $g_2$ are the coupling constants to the cavity mode for the atoms $A$ and $B$, respectively.

When maximum two-photon is considered to be present in the cavity~(i.e., the total excitation is equal to two) and if there is no dissipation in the cavity modes, the closed form of the total state vector at any time may be written as~\cite{pkpgsa,mskgsa}:
\begin{eqnarray}\label{statevector1}
\left|\Psi(t)\right\rangle=C_1\left|e_A,e_B,0\right\rangle+C_2\left|e_A,g_B,1\right\rangle+C_3\left|g_A,e_B,1\right\rangle+C_4\left|g_A,g_B,2\right\rangle,
\end{eqnarray}
where $C_i$~$(i=1,2,3,4)$ are the time-dependent probability amplitudes. For example, $\left|g_A,g_B,2\right\rangle$ represents that the atoms are in their ground states while there are two photons inside the cavity. For this case, the dynamics of the atoms and single mode cavity field can be solved by using time-dependent Schr\"{o}dinger equation and recently many studies have been reported for this case~\cite{pkpgsa,mskgsa,deng}. On the other hand, when cavity decay is taken into account, since the cavity decay only changes the number of photons inside the cavity, the atoms and the cavity mode can be found in any of the states~\cite{mmsg,pkpgsa,smmsz}: $\left|e_A,e_B,0\right\rangle,\left|e_A,g_B,1\right\rangle,\left|e_A,g_B,0\right\rangle,\left|g_A,e_B,1\right\rangle,\left|g_A,e_B,0\right\rangle,\left|g_A,g_B,2\right\rangle,\left|g_A,g_B,1\right\rangle,\left|g_A,g_B,0\right\rangle$. Then the density matrix of the system of two atoms interacting with a single mode cavity field will evolve according to the master equation~\cite{mmsg,pkpgsa,gsasdg}:
\begin{eqnarray}\label{mastereqn}
\dot{\rho}&=&-\frac{i}{\hbar}\left[H,\rho\right]-\frac{\kappa}{2}\left(a^{\dagger}a\rho-2a\rho a^{\dagger}+\rho a^{\dagger}a\right),
\end{eqnarray}
where $\sigma_+^i=\left|e_i\right\rangle\left\langle g_i\right|$ and $\sigma_-^i=\left|g_i\right\rangle\left\langle e_i\right|$~$(i=A,B)$ are the spin inversion operators. Here the first term in Eq.~(\ref{mastereqn}) is generated by the Hamiltonian given in Eq.~(\ref{hamiltonian}) and represents the coherent evolution, while  the other  term denotes the incoherent dynamics; $\kappa$ is the rate of loss of photons from the cavity mode. By setting $\{\left|1\right\rangle\equiv\left|e_A,e_B,0\right\rangle$, $\left|2\right\rangle\equiv\left|e_A,g_B,1\right\rangle$, $\left|3\right\rangle\equiv\left|e_A,g_B,0\right\rangle$, $\left|4\right\rangle\equiv\left|g_A,e_B,1\right\rangle$, $\left|5\right\rangle\equiv\left|g_A,e_B,0\right\rangle$, $\left|6\right\rangle\equiv\left|g_A,g_B,2\right\rangle$, $\left|7\right\rangle\equiv\left|g_A,g_B,1\right\rangle$, $\left|8\right\rangle\equiv\left|g_A,g_B,0\right\rangle\}$ as a basis and considering the Eqs.~(\ref{hamiltonian}) and~(\ref{mastereqn}), the differential equations governing the dynamics of two atoms and the single mode cavity field can be obtained easily. In our analysis we will restrict ourselves for the initial states: $\rho(0)=\left|e_A,e_B,0\right\rangle\left\langle e_A,e_B,0\right|, \rho(0)=\left|e_A,g_B,1\right\rangle\left\langle e_A,g_B,1\right|$ and $\rho(0)=\left|\Psi(0)\right\rangle\left\langle\Psi(0)\right|$, where $\left|\Psi(0)\right\rangle=\frac{1}{\sqrt{2}}(\left|g_A,e_B,1\right\rangle+\left|e_A,g_B,1\right\rangle)$. One should note for these initial states the diagonal elements and some of the off-diagonal elements form a closed evolution. They can be listed as:
\begin{eqnarray}\label{solmaster}
\dot{\rho}_{1,1}&=&ig_1(\rho_{1,4}-\rho_{4,1})+ig_2(\rho_{1,2}-\rho_{2,1}),\nonumber\\
\dot{\rho}_{2,2}&=&i\sqrt{2}g_1(\rho_{2,6}-\rho_{6,2})+ig_2(\rho_{2,1}-\rho_{1,2})-\kappa\rho_{2,2},\nonumber\\
\dot{\rho}_{3,3}&=&ig_1(\rho_{3,7}-\rho_{7,3})+\kappa\rho_{2,2}\nonumber\\
\dot{\rho}_{4,4}&=&ig_1(\rho_{4,1}-\rho_{1,4})+i\sqrt{2}g_2(\rho_{4,6}-\rho_{6,4})-\kappa\rho_{4,4},\nonumber\\
\dot{\rho}_{5,5}&=&ig_2(\rho_{5,7}-\rho_{7,5})+\kappa\rho_{4,4},\nonumber\\
\dot{\rho}_{6,6}&=&i\sqrt{2}g_1(\rho_{6,2}-\rho_{2,6})+i\sqrt{2}g_2(\rho_{6,4}-\rho_{4,6})-2\kappa\rho_{6,6},\nonumber\\
\dot{\rho}_{7,7}&=&ig_1(\rho_{7,3}-\rho_{3,7})+ig_2(\rho_{7,5}-\rho_{5,7})+\kappa(2\rho_{6,6}-\rho_{7,7}),\nonumber\\
\dot{\rho}_{8,8}&=&\kappa\rho_{7,7},\nonumber\\
\dot{\rho}_{1,2}&=&-i\delta_2\rho_{1,2}+ig_1(\sqrt{2}\rho_{1,6}-\rho_{4,2})+ig_2(\rho_{1,1}-\rho_{2,2})-\frac{1}{2}\kappa\rho_{1,2},\nonumber\\
\dot{\rho}_{1,4}&=&-i\delta_1\rho_{1,4}+ig_1(\rho_{1,1}-\rho_{4,4})+ig_2(\sqrt{2}\rho_{1,6}-\rho_{2,4})-\frac{1}{2}\kappa\rho_{1,4},\nonumber\\
\dot{\rho}_{1,6}&=&-i(\delta_1+\delta_2)\rho_{1,6}+ig_1(\sqrt{2}\rho_{1,2}-\rho_{4,6})+ig_2(\sqrt{2}\rho_{1,4}-\rho_{2,6})-\kappa\rho_{1,6},\nonumber\\
\dot{\rho}_{2,4}&=&-i(\delta_1-\delta_2)\rho_{2,4}+ig_1(\rho_{2,1}-\sqrt{2}\rho_{6,4})+ig_2(\sqrt{2}\rho_{2,6}-\rho_{1,4})-\kappa\rho_{2,4},\nonumber\\
\dot{\rho}_{2,6}&=&-i\delta_1\rho_{2,6}+i\sqrt{2}g_1(\rho_{2,2}-\rho_{6,6})+ig_2(\sqrt{2}\rho_{2,4}-\rho_{1,6})-\frac{3}{2}\kappa\rho_{2,6},\nonumber\\
\dot{\rho}_{3,5}&=&-i(\delta_1-\delta_2)\rho_{3,5}-ig_1\rho_{7,5}+ig_2\rho_{3,7}+\kappa\rho_{2,4},\nonumber\\
\dot{\rho}_{3,7}&=&-i\delta_1\rho_{3,7}+ig_1(\rho_{3,3}-\rho_{7,7})+ig_2\rho_{3,5}+\frac{\kappa}{2}(2\sqrt{2}\rho_{2,6}-\rho_{3,7}),\nonumber\\
\dot{\rho}_{4,6}&=&-i\delta_2\rho_{4,6}+ig_1(\sqrt{2}\rho_{4,2}-\rho_{1,6})+i\sqrt{2}g_2(\rho_{4,4}-\rho_{6,6})-\frac{3}{2}\kappa\rho_{4,6},\nonumber\\
\dot{\rho}_{5,7}&=&-i\delta_2\rho_{5,7}+ig_1\rho_{5,3}+ig_2(\rho_{5,5}-\rho_{7,7})+\frac{\kappa}{2}(2\sqrt{2}\rho_{4,6}-\rho_{5,7}),
\end{eqnarray}
where $\delta_1=\omega_A-\omega$ and $\delta_2=\omega_B-\omega$ are called the detunings and describes the frequency difference between the energy levels of atoms and the cavity mode, $\rho_{i,j}=\left\langle i\right|\rho\left|j\right\rangle$ and $\rho_{i,j}=\rho_{j,i}^*$.  We will consider two particular detuning types: for $\delta_1=\delta_1=\delta$ the atoms will be called identical since the transition frequencies of the atoms are the same and the atoms will be called unidentical for $\delta_1=-\delta_1=\delta$ case for which the individual atom transition frequencies are different~\cite{mskgsa}. In practice, the detunings might be slightly different due to broadening, in this work we will consider relatively large values of $\delta$.  One should note that it is quite difficult to obtain an analytic solution for this set of coupled differential equations, therefore we will solve these equations numerically for the considered initial states.

The density matrix of the two atoms is obtained by taking a partial trace over the cavity degrees of freedom: $\rho_{AB}(t)=Tr_c\{\rho\}=\sum_{n=0}^2\left\langle n\right|\rho\left|n\right\rangle$ and used to calculate the quantum correlations as measured by CHSH-Bell inequality, concurrence and quantum discord between two atoms interacting with a single mode cavity field. For the considered initial states whose dynamics is given by Eq.~(\ref{solmaster}), the  reduced density matrix of the atoms in the two-qubit standart basis $\{\left|1\right\rangle\equiv\left|e_A,e_B\right\rangle,\left|2\right\rangle\equiv\left|e_A,g_B\right\rangle,\left|3\right\rangle\equiv\left|g_A,e_B\right\rangle,\left|4\right\rangle\equiv\left|g_A,g_B\right\rangle\}$  can be calculated as:
\begin{eqnarray}\label{reducedden}
\rho_{AB}(t)=
\left(\begin{array}[pos]{cccc} a & 0 & 0 & 0 \\ 0 & b & e & 0 \\ o & e^* & c & 0 \\ 0 & 0 & 0 & d \end{array}\right)
\end{eqnarray}
where $a=\rho_{1,1}, b=\rho_{2,2}+\rho_{3,3}, c=\rho_{4,4}+\rho_{5,5}, d=\rho_{6,6}+\rho_{7,7}+\rho_{8,8}$ and $e=\rho_{2,4}+\rho_{3,5}$. It should be noted that the density matrix in Eq.~(\ref{reducedden}) has an X-form and the dynamics preserve its form. The considered correlation measures for this type of states can be calculated easily~\cite{ara,fare1,hms,wlnl,mbfc}. The concurrence~\cite{wootters} as an entanglement measure is given by~\cite{fare1}:
\begin{eqnarray}\label{conc}
C(t)=2\max\{0,|e|-\sqrt{ad}\}.
\end{eqnarray}
The Bell nonlocality measured by the CHSH-inequality~\cite{chsh} can be found as~\cite{mbfc}:
\begin{eqnarray}\label{bell}
\mathcal{B}(t)=\max\{\mathcal{B}_1(t),\mathcal{B}_2(t)\},\nonumber\\
\mathcal{B}_1(t)=2\sqrt{u1+u2},\quad \mathcal{B}_2(t)=2\sqrt{u1+u3},
\end{eqnarray}
where $u1=u3=4|e|^2$ and $u2=(a+d-b-c)^2$.

QD is defined as~\cite{howhz}:
\begin{eqnarray}\label{qd1}
D(t)=I(t)-J(t),
\end{eqnarray}
where $I(t)=S(\rho_A)+S(\rho_B)-S(\rho_{AB})$ measures the total correlation between the atoms; $S(\rho)=-Tr(\rho\log_2\rho)$ is the von-Neumann entropy and $\rho_A~(\rho_B)$ is the reduced density matrix  obtained by tracing $\rho_{AB}$ over the subsystem $B~(A)$. The other quantity $J(t)$ is the amount of classical correlations between the atoms defined as the maximum information one can get about the atom $A$ by performing  measurements on the atom $B$, or vice versa. Its calculation requires maximization over one of the subsystem and generally obtaining an analytic result is not an easy task~\cite{ara}. Recently, for the density matrix having X-form the explicit expressions of QD and classical correlation are reported~\cite{ara,hms,wlnl}. We have used the expressions given in Ref.~\cite{wlnl}. The classical correlation $J(t)$ for the density matrix Eq.~(\ref{reducedden}) is given as:
\begin{eqnarray}\label{class}
J(t)=\max\{C_1,C_2\},
\end{eqnarray}
where $C_j=M(a+b)-P_j$. And QD is given as:
\begin{eqnarray}\label{qd2}
D(t)=\min\{Q_1,Q_2\},
\end{eqnarray}
where $Q_j=M(a+c)+\sum_{i=1}^4\lambda_{AB}^i\log_2\lambda_{AB}^i+P_j$, with $\lambda_{AB}^j$ being the eigenvalues of $\rho_{AB}$, $P_1=M(\tau)$, $P_2=-(a\log_2a+b\log_2b+c\log_2c+d\log_2d)-M(a+c)$, $\tau=(1+\sqrt{(1-2(c+d))^2+4|e|^2})/2$ and $M(\alpha)=-\alpha\log_2\alpha-(1-\alpha)\log_2(1-\alpha)$. We have also calculated QD numerically by considering projective measurements~\cite{howhz,wsfb} and observed that the results obtained by using Eq.~(\ref{qd2}) and the numerical results agree perfectly. 

As is well known, concurrence is equal to 1~(0) for a maximally entangled~(separable) state. Although QD is equal to 1 for a maximally entangled state, it may or may not be equal to zero for a separable state because it was shown that even some separable states can carry non-zero quantum discord~\cite{ara,lhvv}. For pure states QD is found to be equal to the entanglement of formation~\cite{ara}. On the other hand when $\mathcal{B}(t)>2$ the CHSH-Bell inequality is violated which signifies that the correlations cannot be accessible by any classical local model~\cite{werner}.

\section{Results}
\subsection{Effects of Cavity Damping-Separable States}\label{cavity1}
In this section, we will examine the effect of cavity decay rate, $\kappa$, on the quantum correlations as measured by concurrence, CHSH-Bell inequality and quantum discord. To do this we will solve the differential equations in Eq.~(\ref{solmaster}) numerically for the initial states: $\rho(0)=\left|e_A,e_B,0\right\rangle\left\langle e_A,e_B,0\right|, \rho(0)=\left|e_A,g_B,1\right\rangle\left\langle e_A,g_B,1\right|$ and  $\rho(0)=\left|\Psi(0)\right\rangle\left\langle\Psi(0)\right|$, where $\left|\Psi(0)\right\rangle=\frac{1}{\sqrt{2}}(\left|g_A,e_B,1\right\rangle+\left|e_A,g_B,1\right\rangle)$ and we will set $g_1=g_2=g$~(symmetric coupling), $\delta_1=\delta_2=5g$ for identical atoms and $\delta_1=-\delta_2=5g$ for unidentical atoms. One should note that the detunings are chosen somewhat large because at higher detunings the qubit-cavity energy exchange is low~\cite{pkpgsa,mskgsa,fmpp} and a high-degree of cavity-induced quantum correlations can be created~\cite{fmpp}.
\begin{figure}[!ht]\centering
\includegraphics[width=7.5cm]{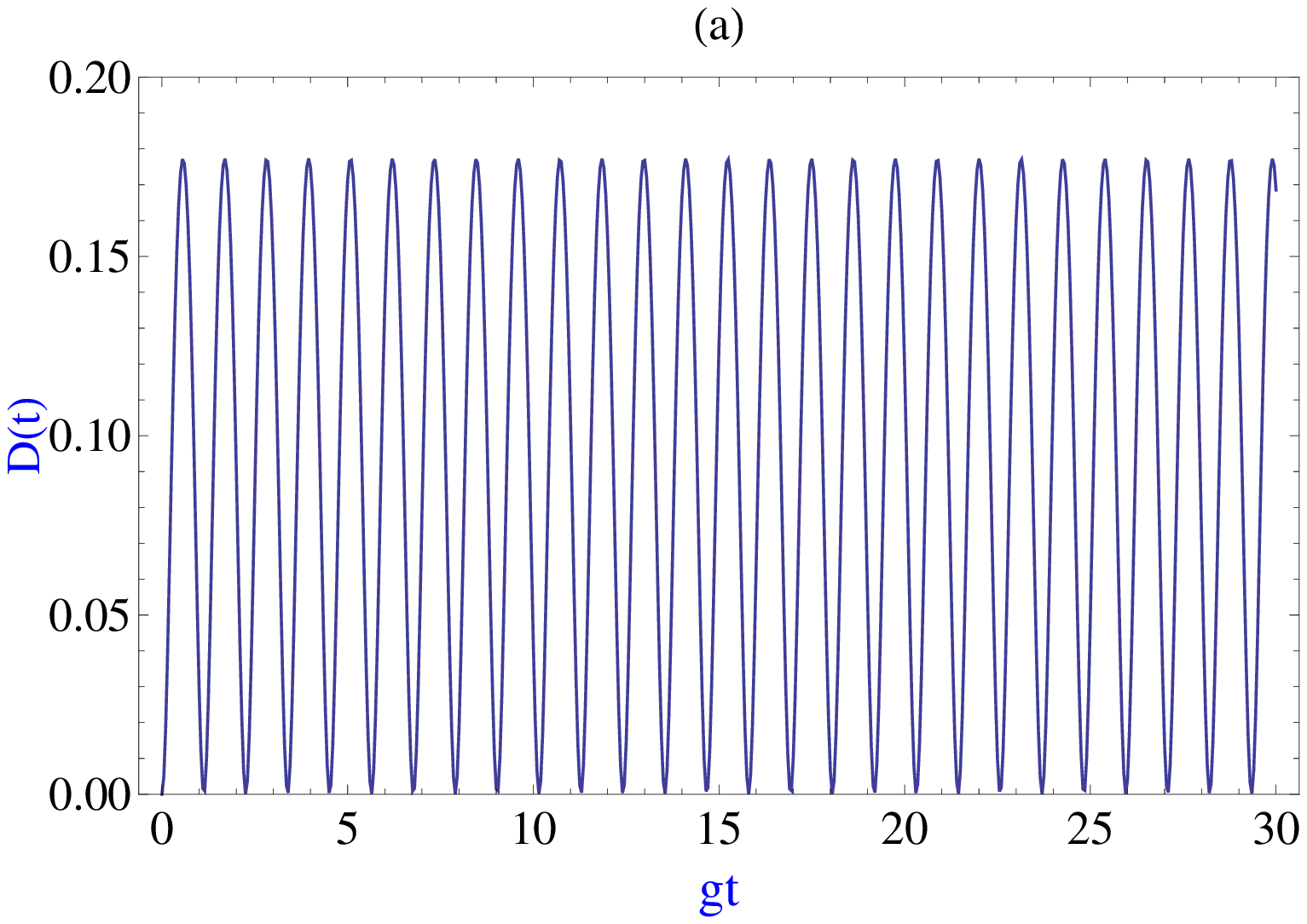}
\includegraphics[width=7.5cm]{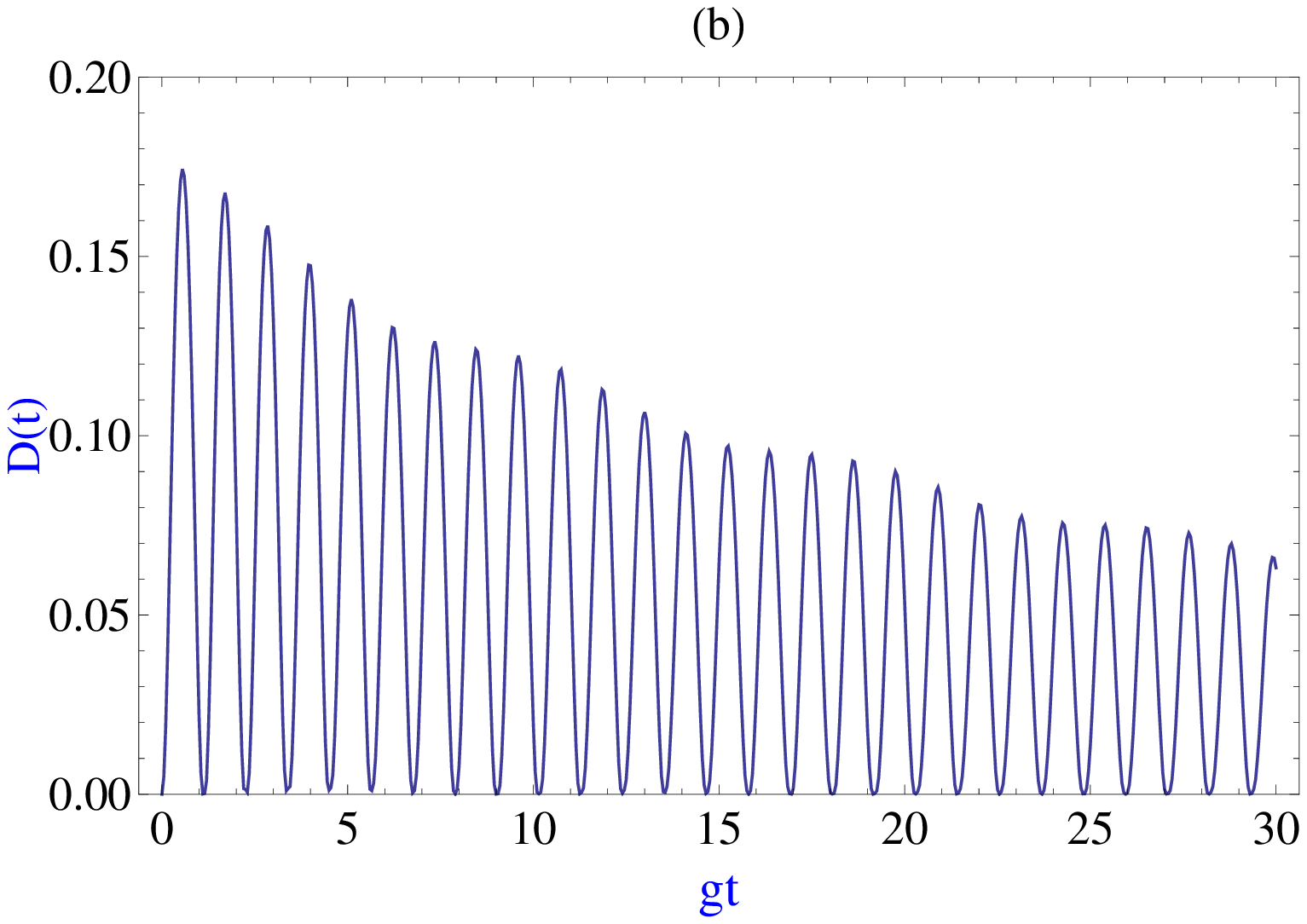}

\includegraphics[width=7.5cm]{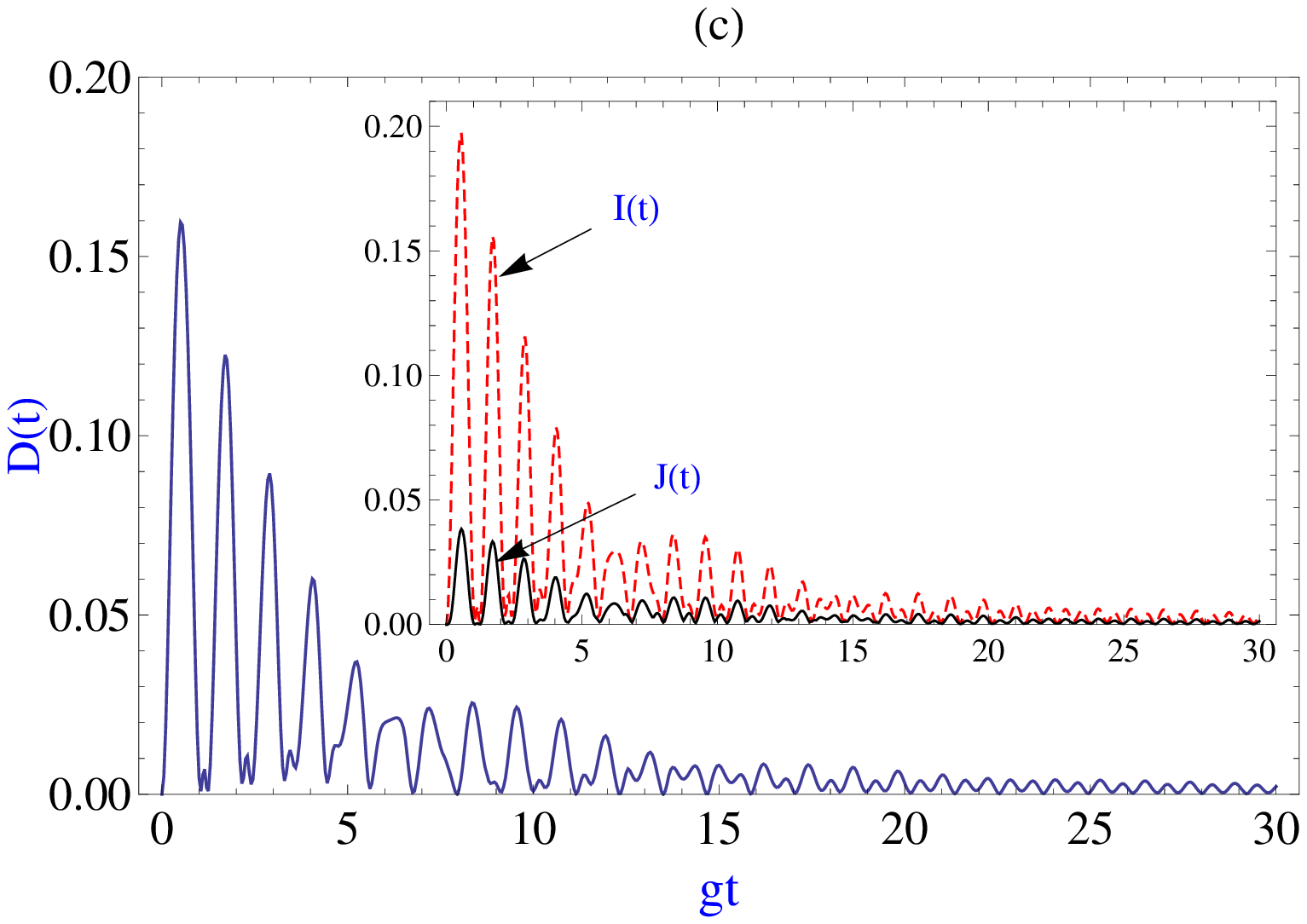}
\includegraphics[width=7.5cm]{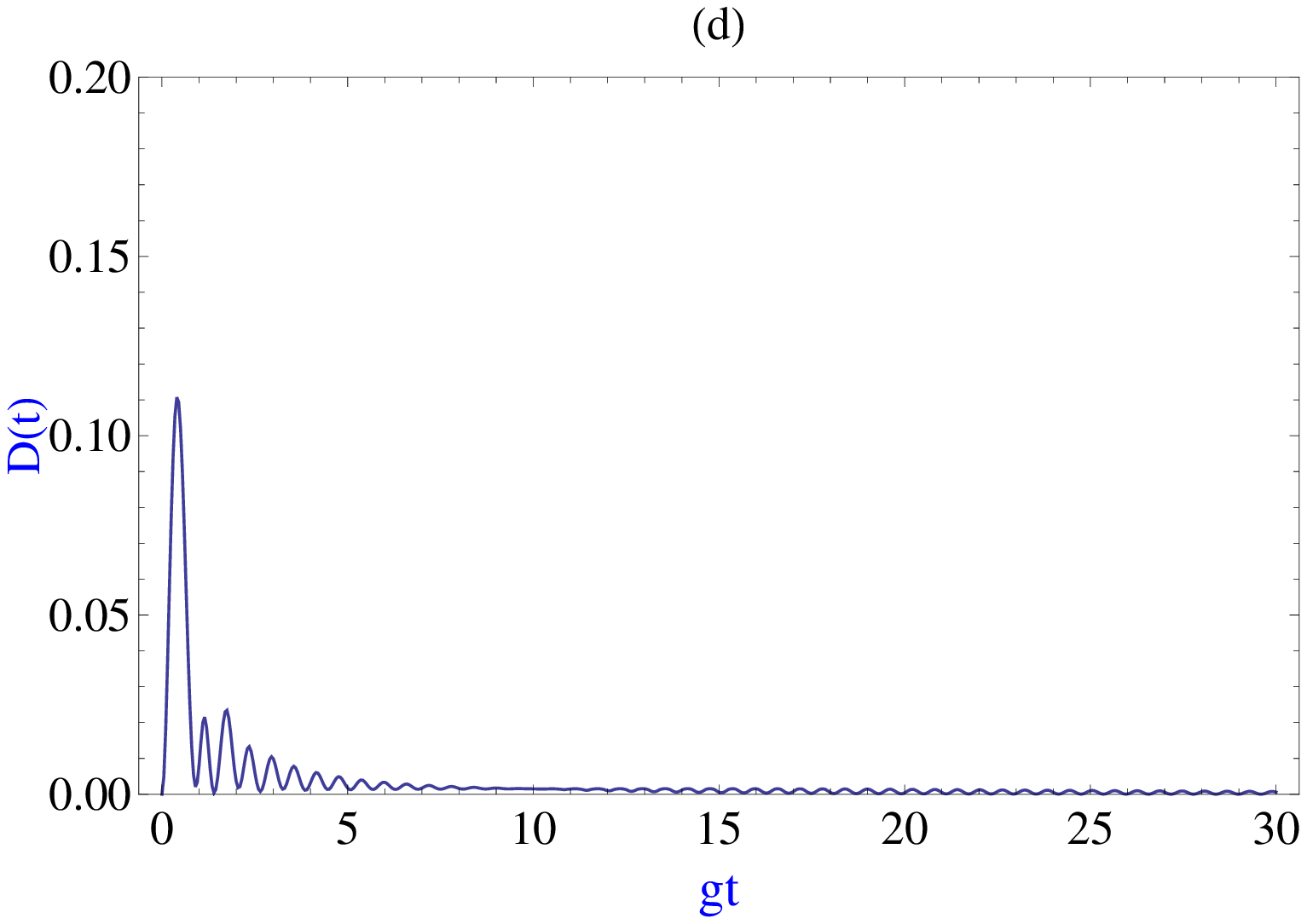}

\includegraphics[width=7.5cm]{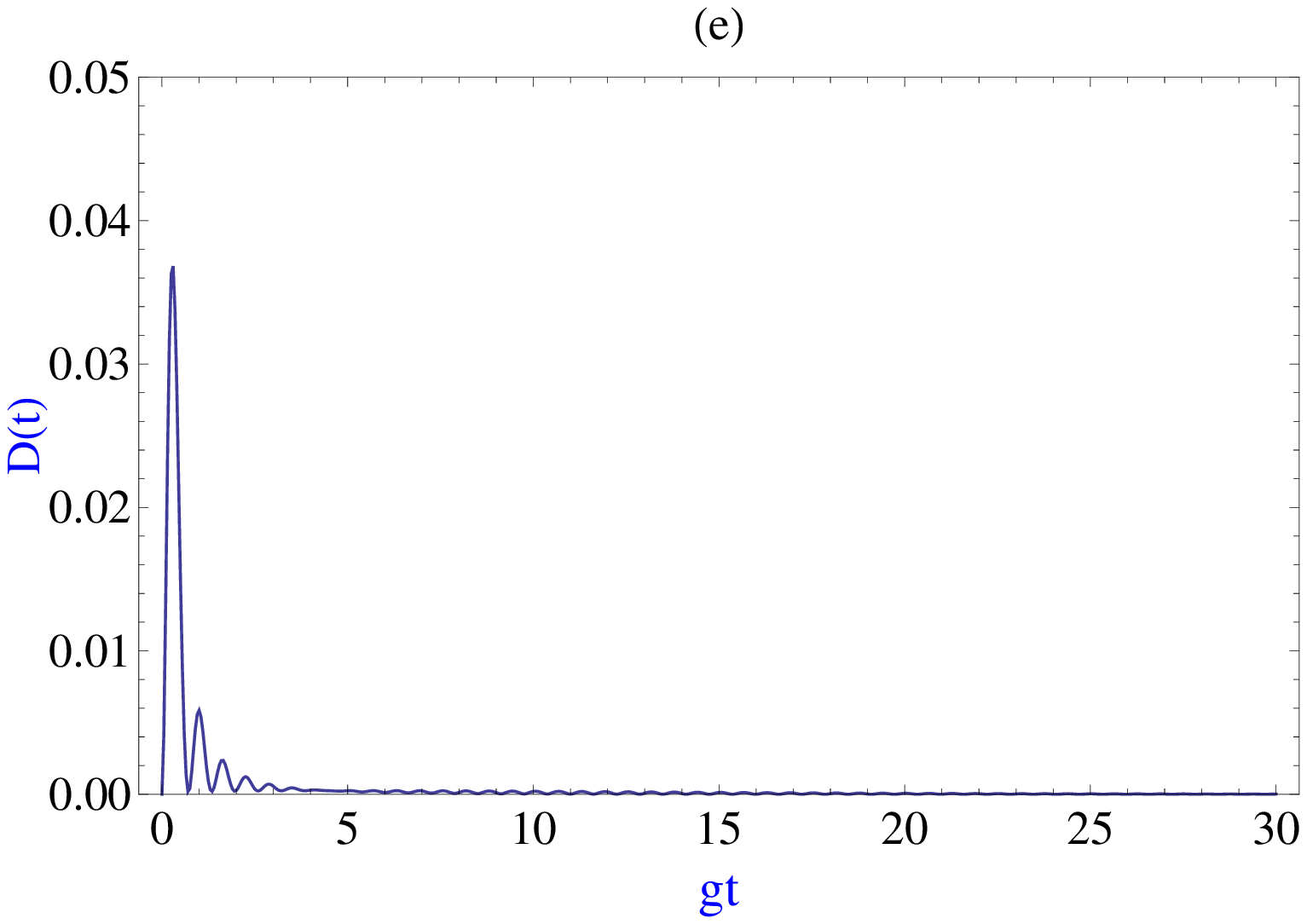}
\caption{\label{fig.1} Quantum discord versus $gt$ for unidentical atoms~($\delta_1=-\delta_2=5g$) and $\rho(0)=\left|e_A,e_B,0\right\rangle\left\langle e_A,e_B,0\right|$ initial state for the decay rates: (a)~$\kappa=0$, (b)~$\kappa=0.02g$, (c)~$\kappa=0.2g$, (d)~$\kappa=2.0g$ and (e)~$\kappa=20g$. Note that under these conditions entanglement and Bell nonlocality are not induced, thus these are not plotted here and the inset in~(c) represents the total~$I(t)$ and classical~$J(t)$ correlations versus $gt$ for $\kappa=0.2g$.}
\end{figure}

We first consider the dynamics of concurrence and quantum discord for initially separable state $\rho(0)=\left|e_A,e_B,0\right\rangle\left\langle e_A,e_B,0\right|$ and display $C$ and $D$ versus dimensionless time, $gt$ in Figs.~1 and~2 for unidentical and identical atoms, respectively. For this initial state neither non-zero concurrence nor violation in Bell inequality ever occurs for unidentical qubits, so their dynamics are not plotted in Fig.~1. However, this does not indicate the loss of quantum correlations; QD induced by the atom-field interaction oscillates nearly between 0 and 0.18 for $\kappa=0$~(Fig.~1(a)). A similar result is found by Sun {\it et al.,} in Ref.~\cite{sls} where the authors studied the entanglement and quantum discord between two qubits which are independently coupled to a common environment modelled as an Ising spin chain. It can be seen from the analysis of Figs.~1(b)-1(e) that QD is created independently of the cavity decay rate, but its maximum value depends on the decay rate $\kappa$ inversely and the effect of cavity decay on the dynamics of QD is to create a damped oscillatory behaviour; damping in QD is proportional to the cavity decay rate. Also, in order to understand the nature of non-zero quantum discord, we have plotted the total correlation, $I(t)$, and classical correlation, $J(t)$, for $\kappa=0.2g$ in the inset of Fig.~1(c). It should be noted that for this initial state, the total as well as classical correlations are also induced and $I(t)$ is always greater than $J(t)$ as a consequence non-zero quantum discord is present. This inset also demonstrates that $I(t)$ and $J(t)$ are also damped oscillatorily as $D(t)$. Correlations in the system are gradually lost to the reservoir because of the cavity decay as found in the system studied in Ref.~\cite{gglxg}.
\begin{figure}[!ht]\centering
\includegraphics[width=7.5cm]{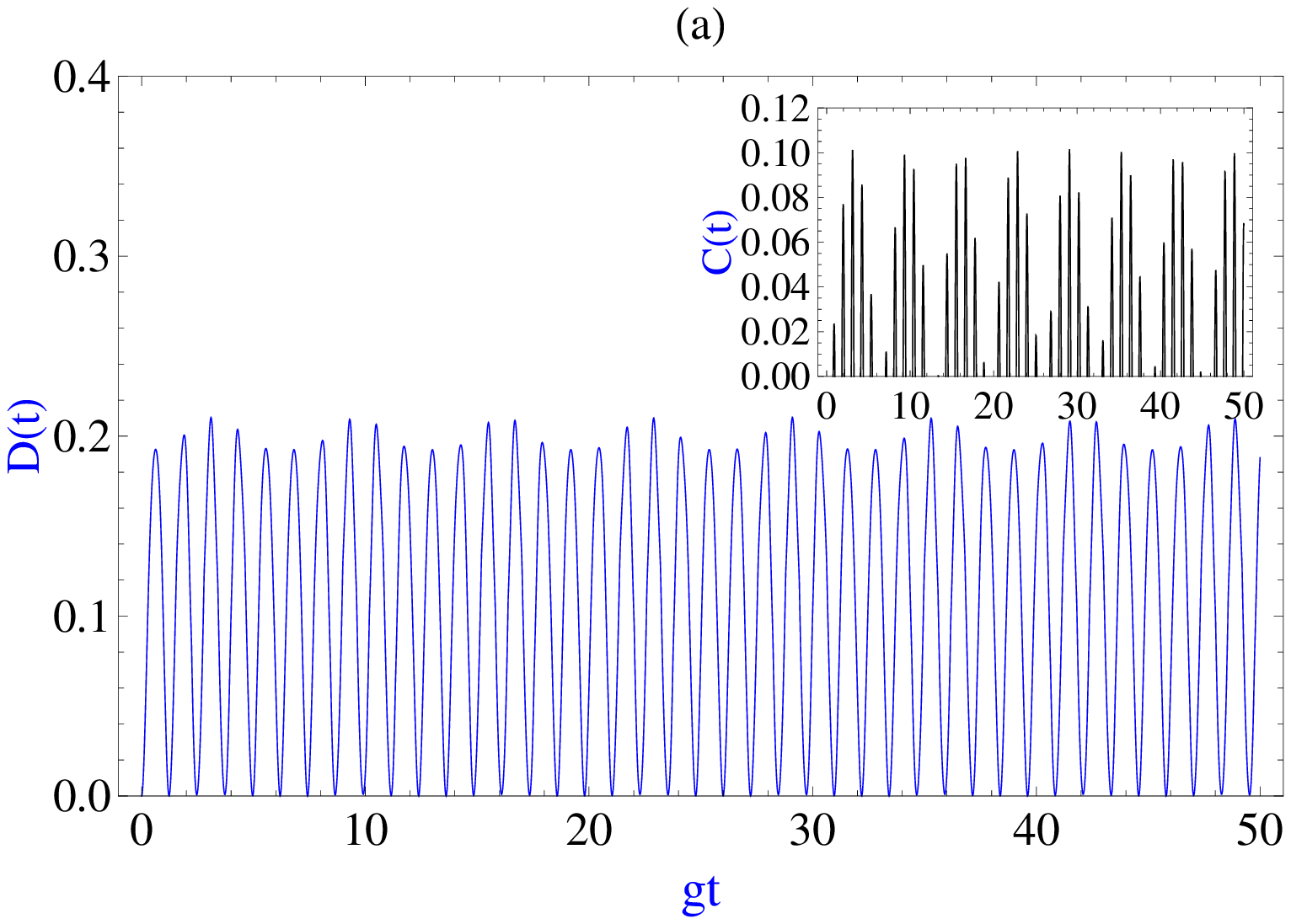}
\includegraphics[width=7.5cm]{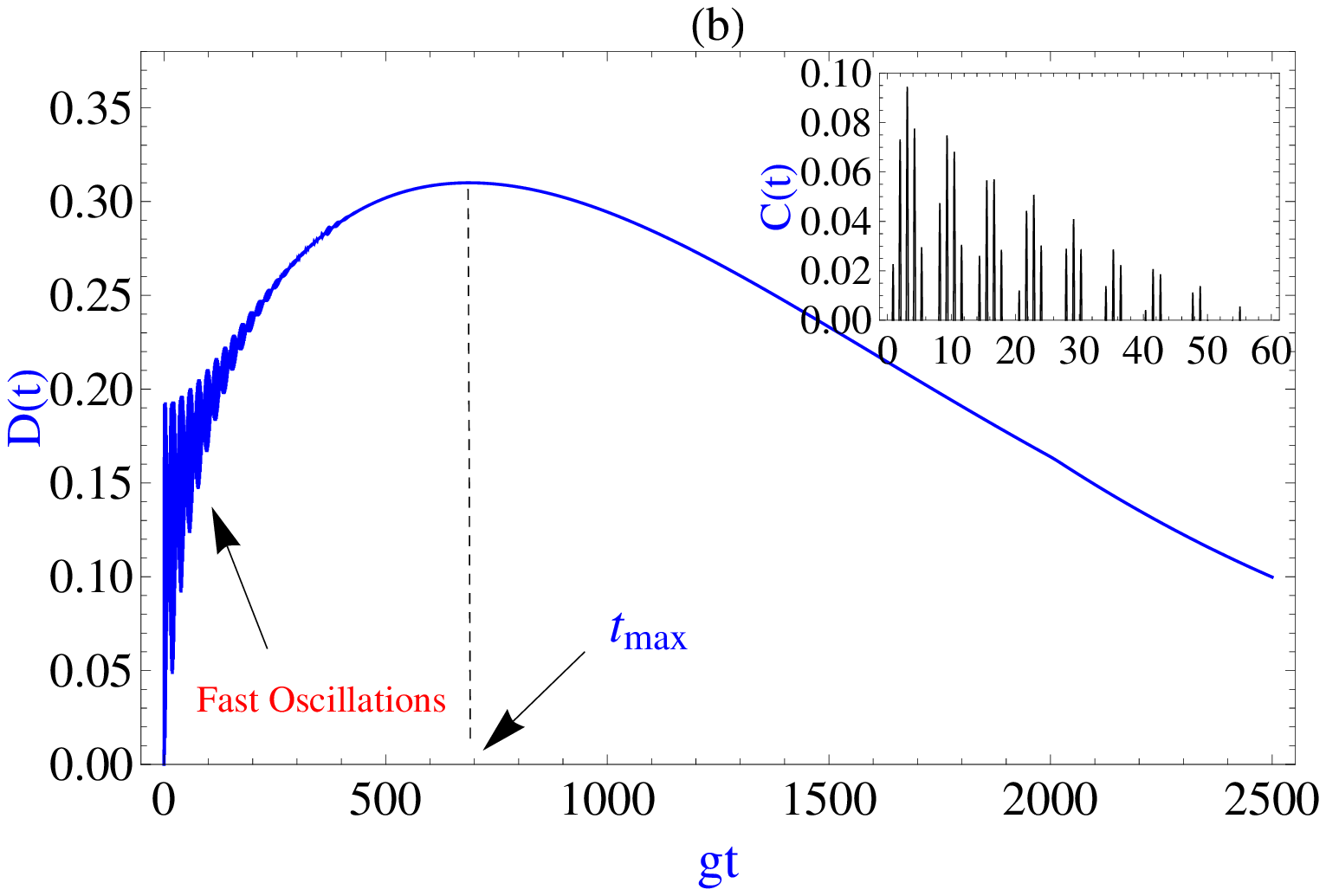}

\includegraphics[width=7.5cm]{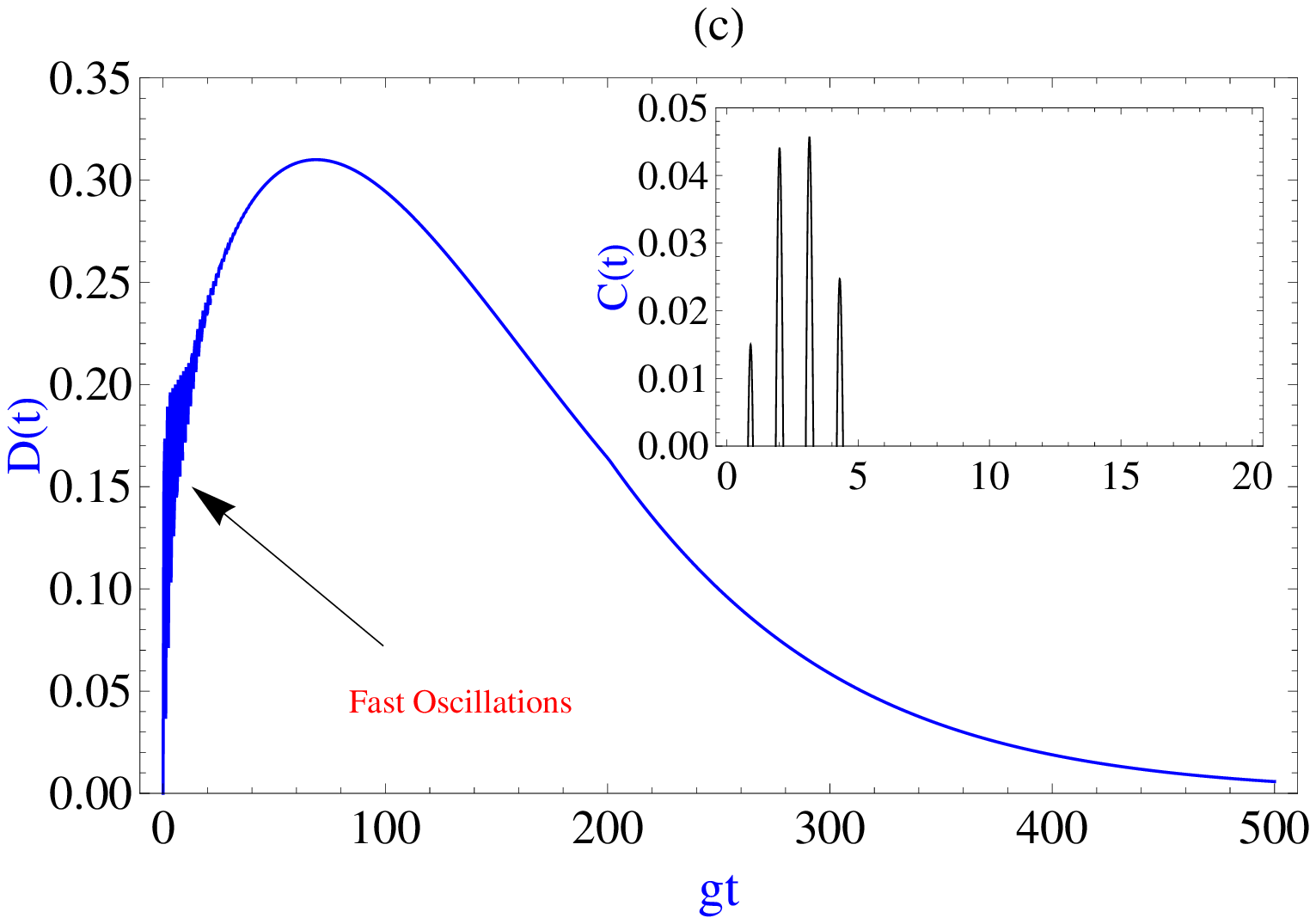}
\includegraphics[width=7.5cm]{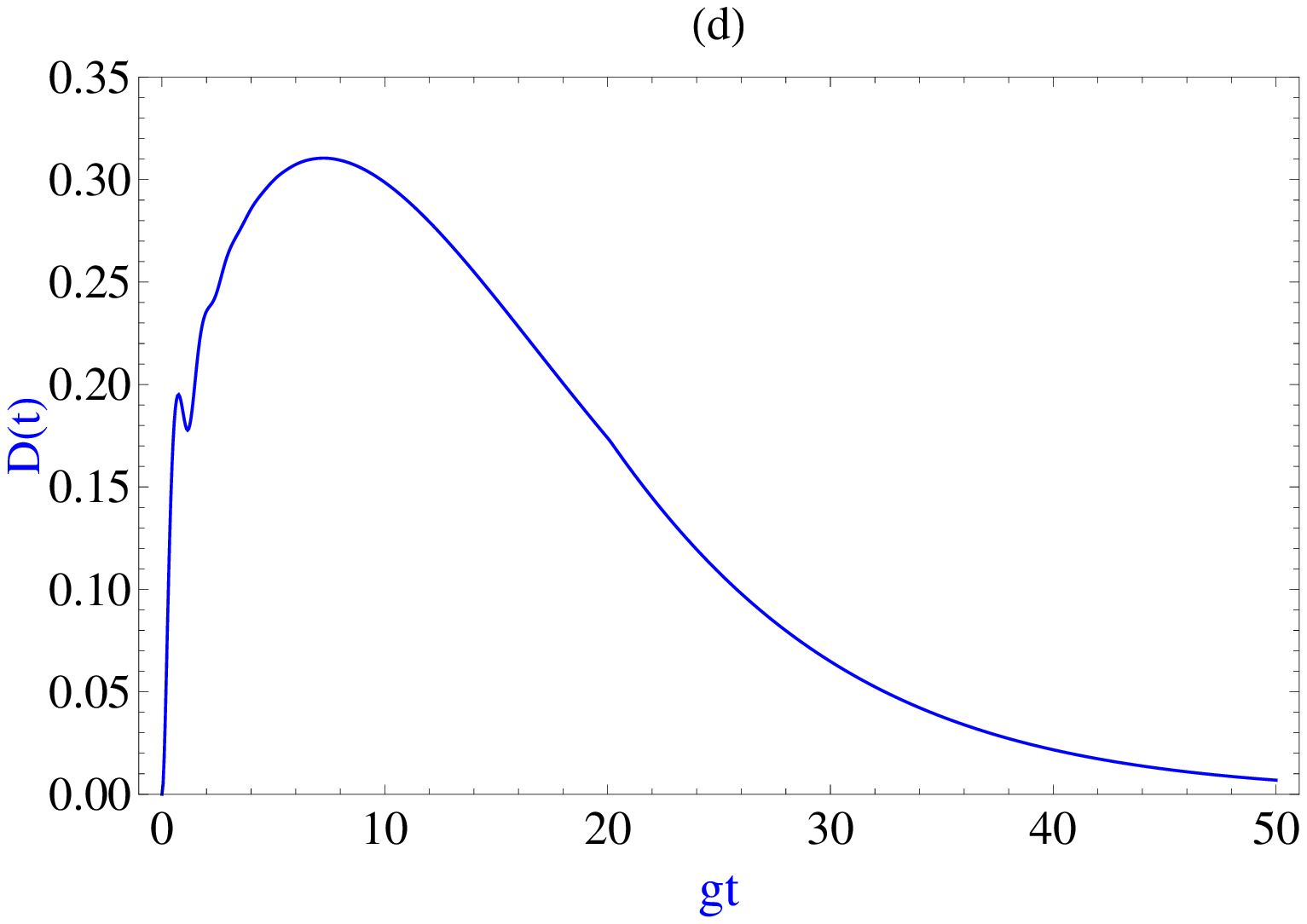}

\includegraphics[width=7.5cm]{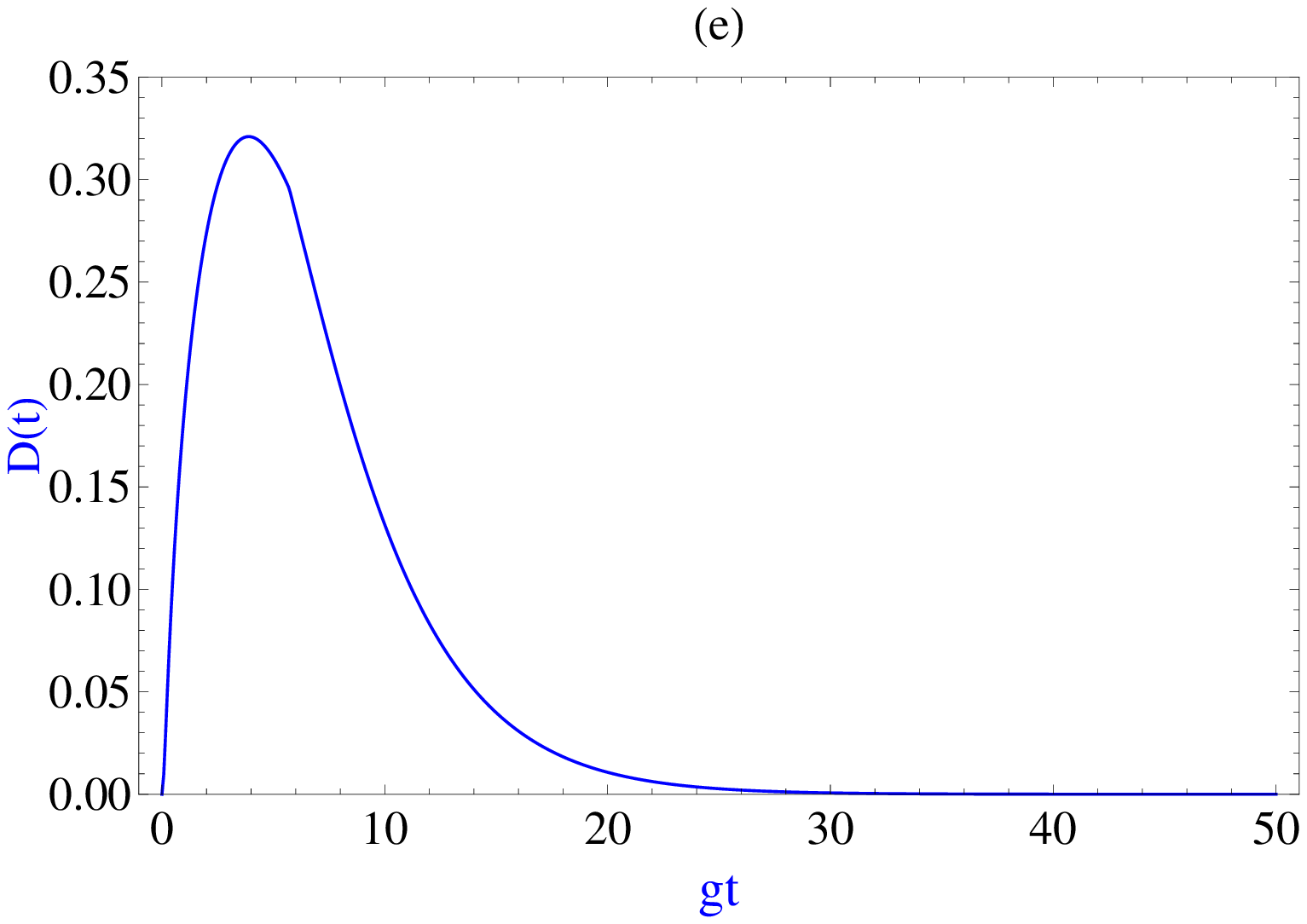}
\caption{\label{fig.2}Quantum discord and concurrence~(insets in~(a),~(b) and~(c)) versus $gt$ for identical atoms~($\delta_1=\delta_2=5g$) and $\rho(0)=\left|e_A,e_B,0\right\rangle\left\langle e_A,e_B,0\right|$ initial state for the decay rates: (a)~$\kappa=0$, (b)~$\kappa=0.02g$, (c)~$\kappa=0.2g$, (d)~$\kappa=2.0g$ and (e)~$\kappa=20g$. Note that under these conditions  Bell nonlocality is not induced, and for $\kappa=2g$ and $\kappa=20g$, $C(t)$ has no dynamics as well, thus they are  not plotted here. The insets are plotted under the same conditions as $D(t)$.}
\end{figure}

We have calculated the dynamics of CHSH inequality, concurrence and quantum discord for the same initial state $\rho(0)=\left|e_A,e_B,0\right\rangle\left\langle e_A,e_B,0\right|$ for the identical atoms~($\delta_1=\delta_2=5g$) case to understand the effect of the type of detuning for the same leaky cavity and plot the resulting $D$ and $C$ as a function of $gt$ in Figs.~2(a)-2(e) and insets of Figs.~2(a)-2(c), respectively.  As can be seen from the insets of Figs.~2(a)-2(c), non-zero entanglement, although small, is created which decays to zero as cavity decay rate is increased and the induced entanglement has no dynamics for higher values of $\kappa$~($\kappa=2g$ and $\kappa=20g$). Bell inequality is found to be not violated for any value of $\kappa$, which is expected because entanglement is low and generally Bell inequality violation is shown to be a hallmark of high $C(t)$~\cite{fare2,bfc,bfc2}, so $\mathcal{B}(t)$ is not displayed in Fig.~2. One should note the difference in the dynamics of QD in the leaky cavity for unidentical and identical atom cases~(Figs.~1 and~2). While QD has an oscillatory decay type dynamics for the unidentical atoms~(Fig.~1), it increases until the time $t=t_{max}$ and then decays for the identical atoms. Also, for the same decay parameters, QD is found to have a much longer lifetime for the identical atoms. The higher leakage in the cavity decreases $t_{max}$ and the lifetime of QD after the time $t=t_{max}$. On the other hand, $\kappa$ has no appreciable effect on the maximum value of $D(t)$. Also note that the fast oscillations in QD are gradually smoothed for larger values of $\kappa$ and longer times as can be seen from Figs.~2(b)-2(e).
\begin{figure}[!ht]\centering
\includegraphics[width=5.1cm]{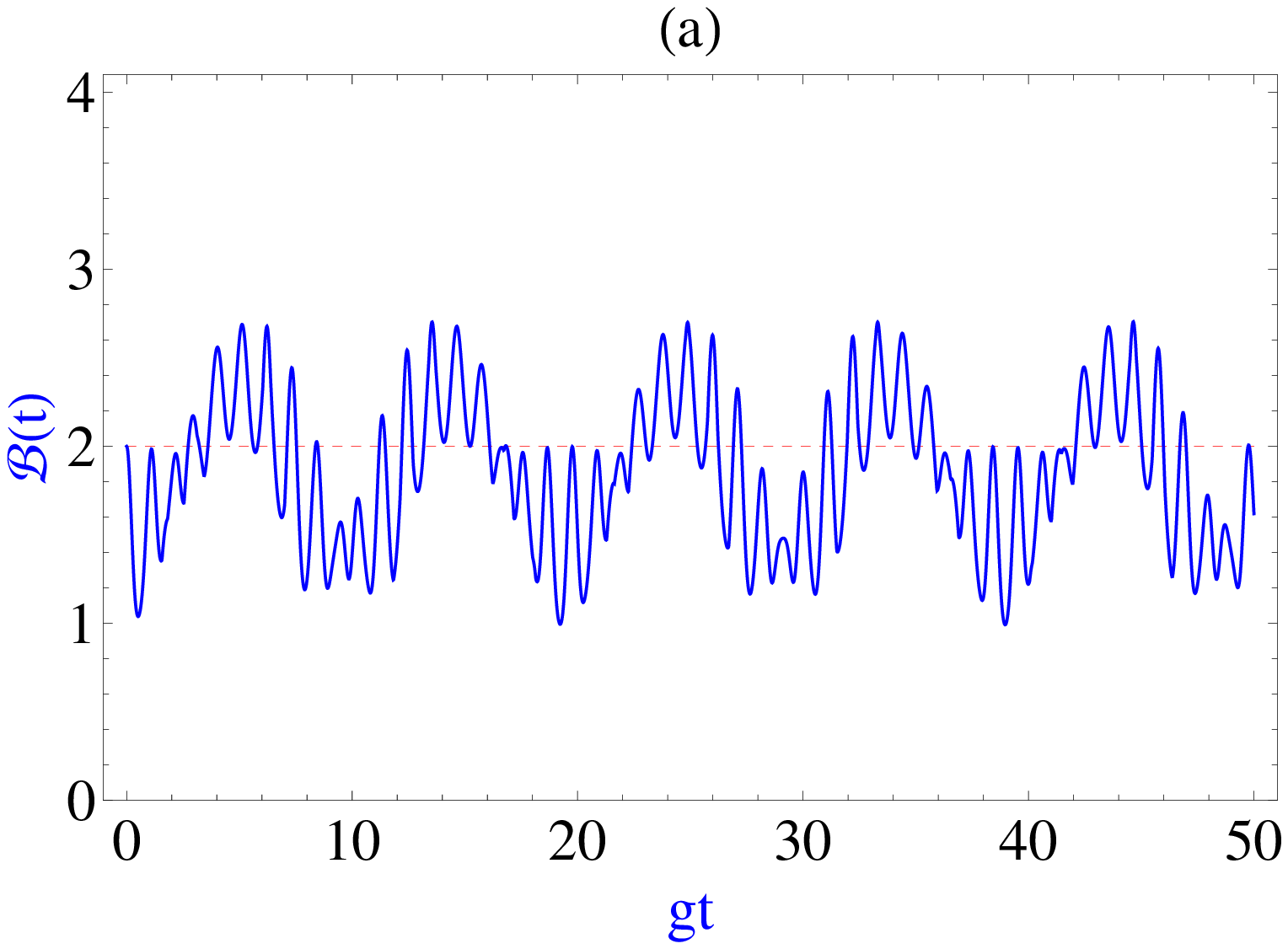}
\includegraphics[width=5.1cm]{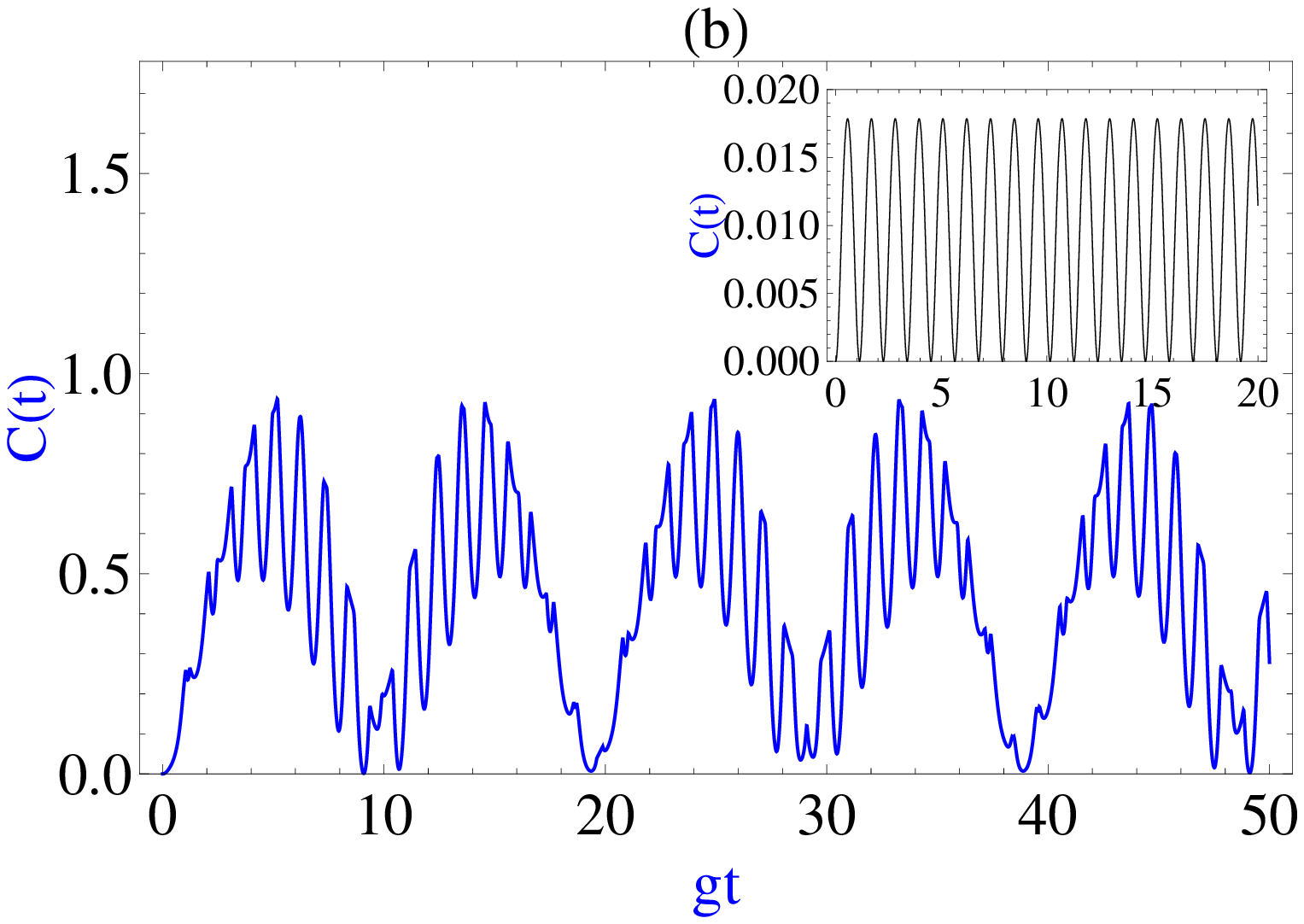}
\includegraphics[width=5.1cm]{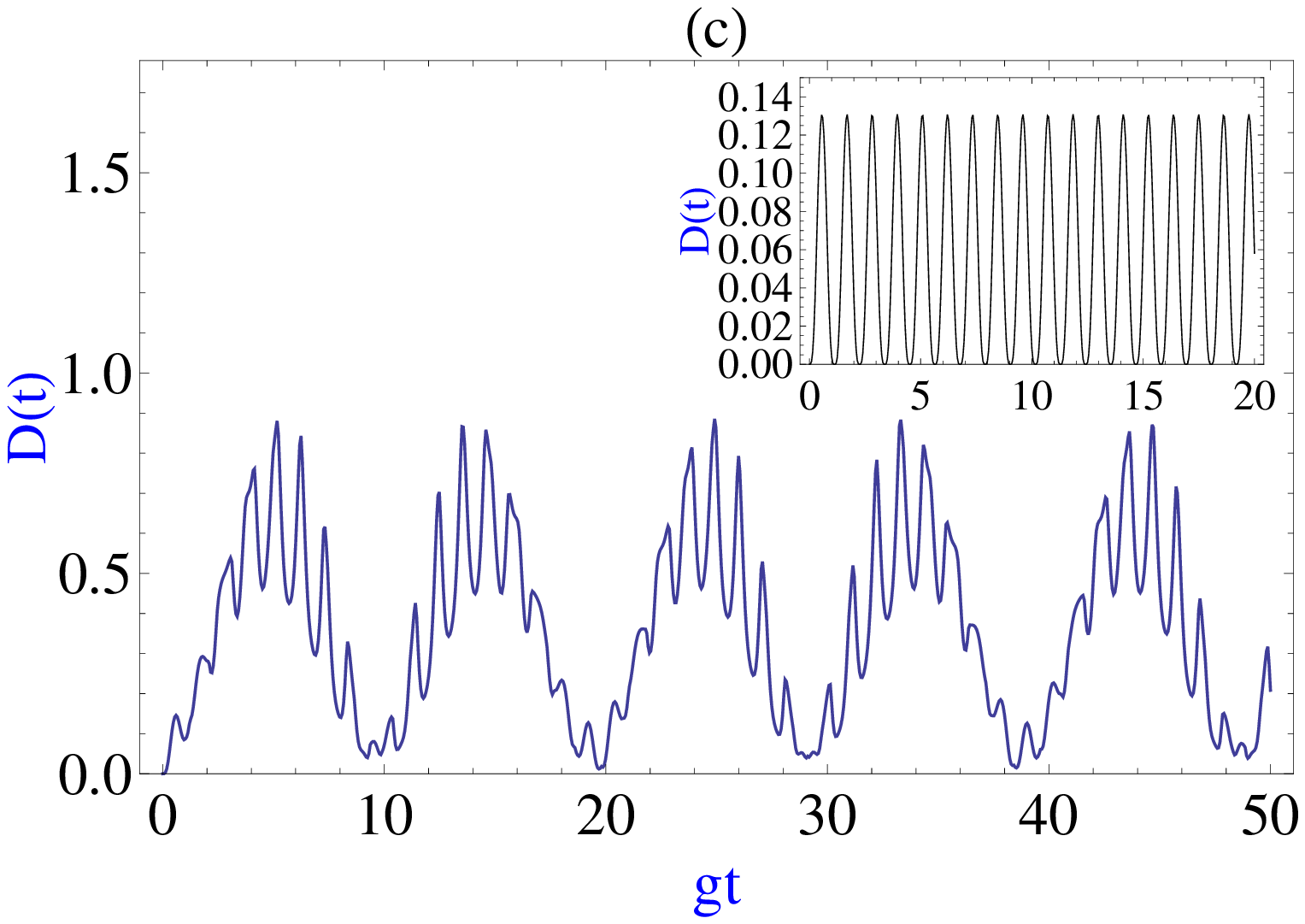}

\includegraphics[width=5.1cm]{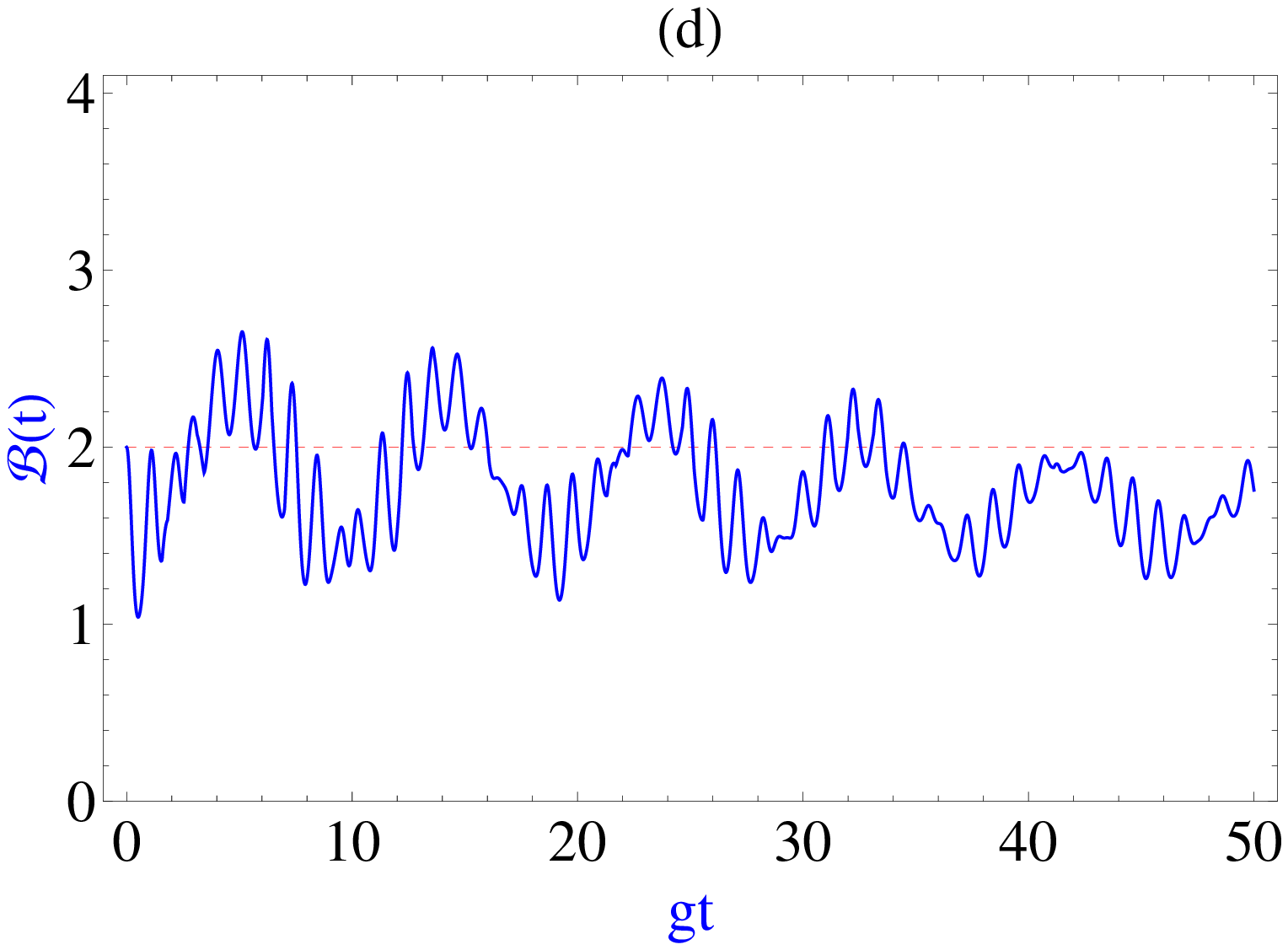}
\includegraphics[width=5.1cm]{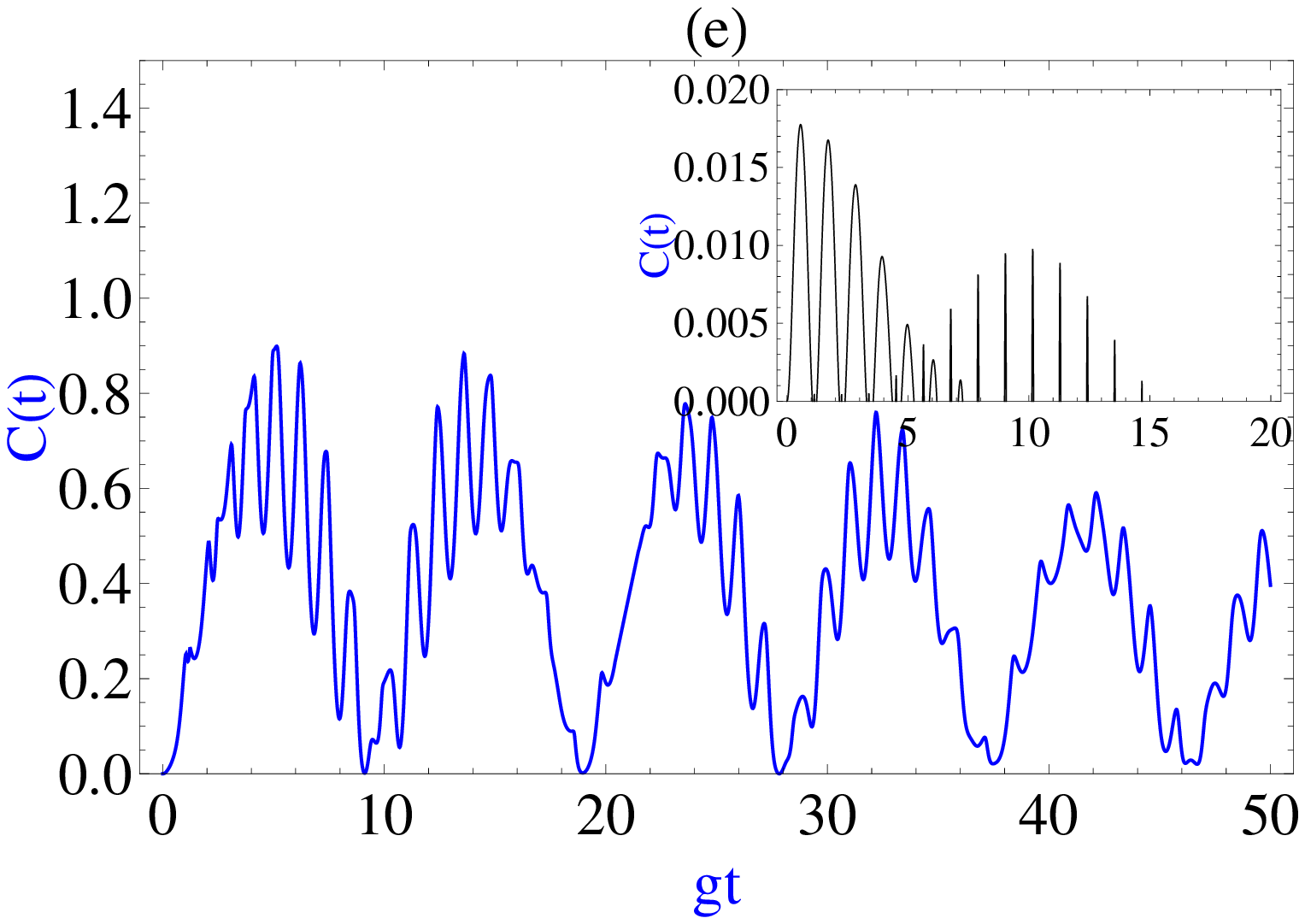}
\includegraphics[width=5.1cm]{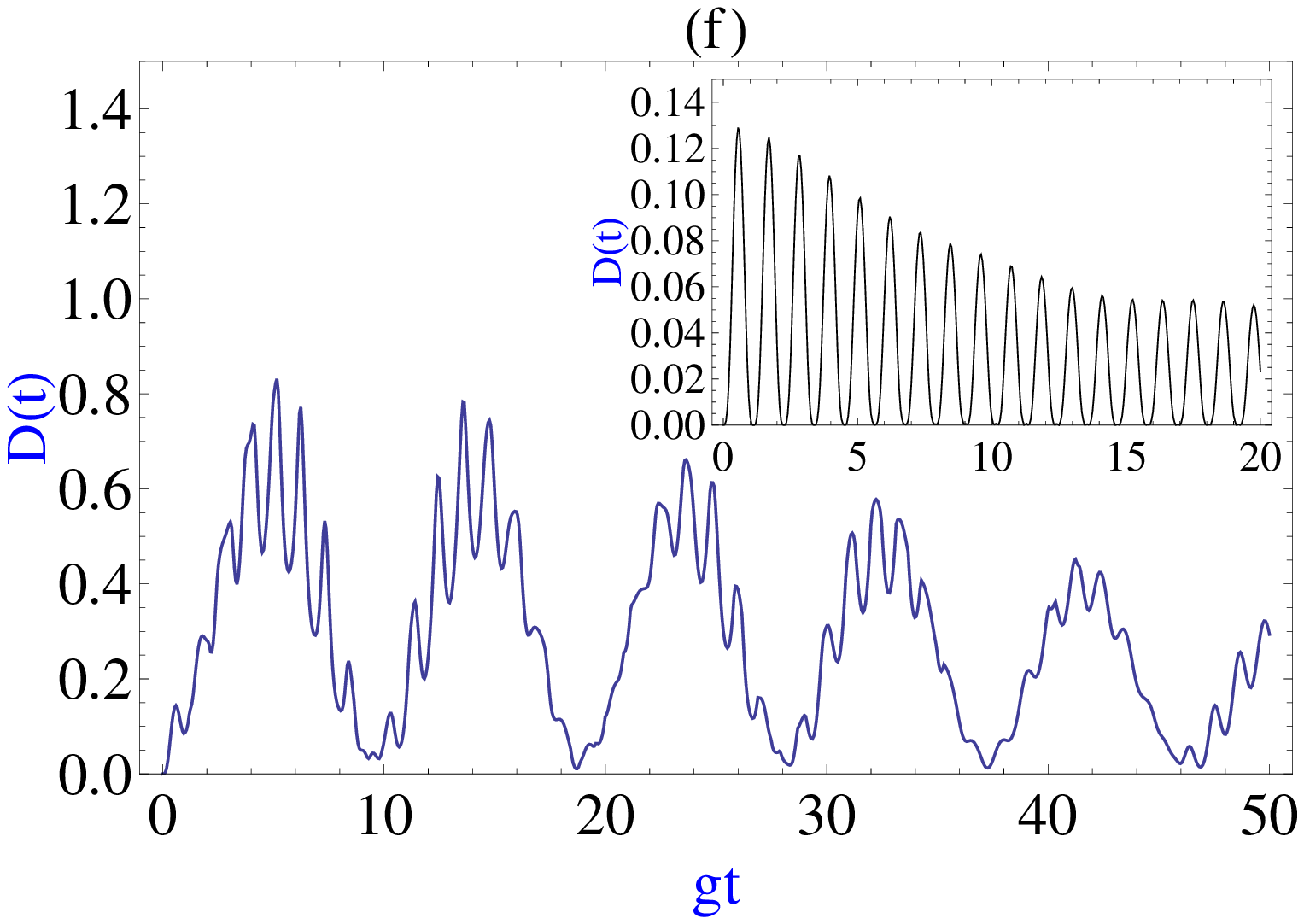}

\includegraphics[width=5.1cm]{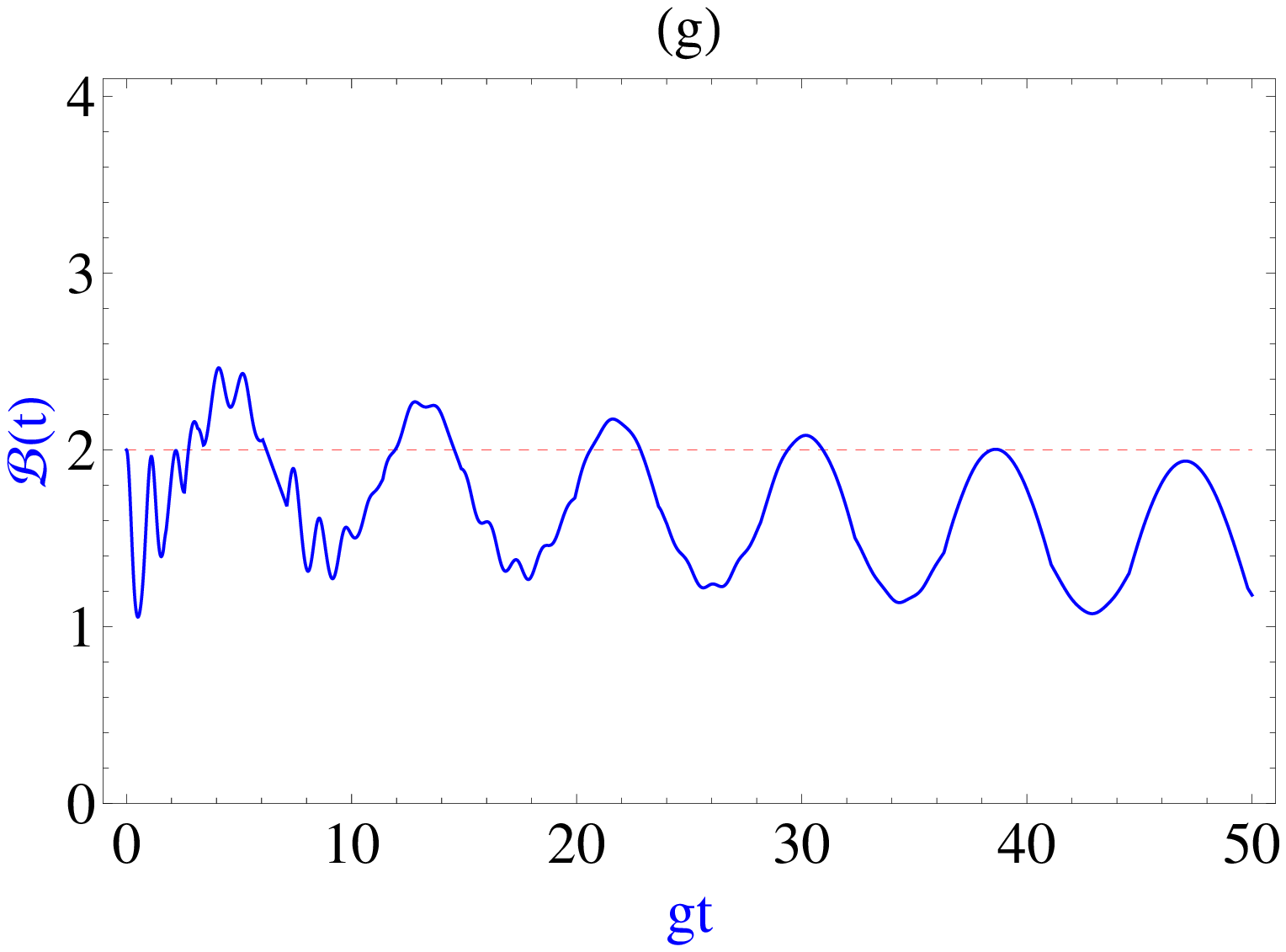}
\includegraphics[width=5.1cm]{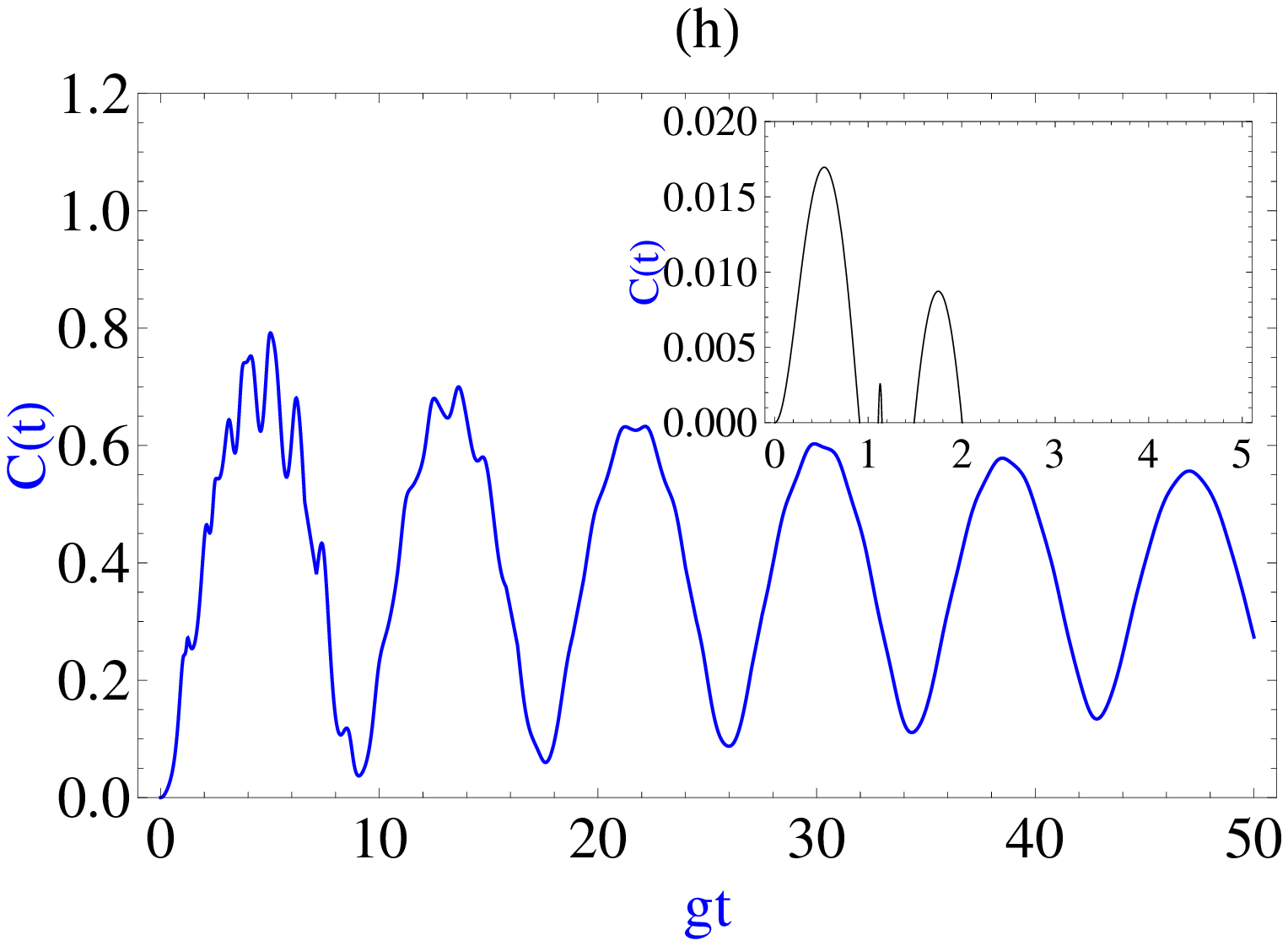}
\includegraphics[width=5.1cm]{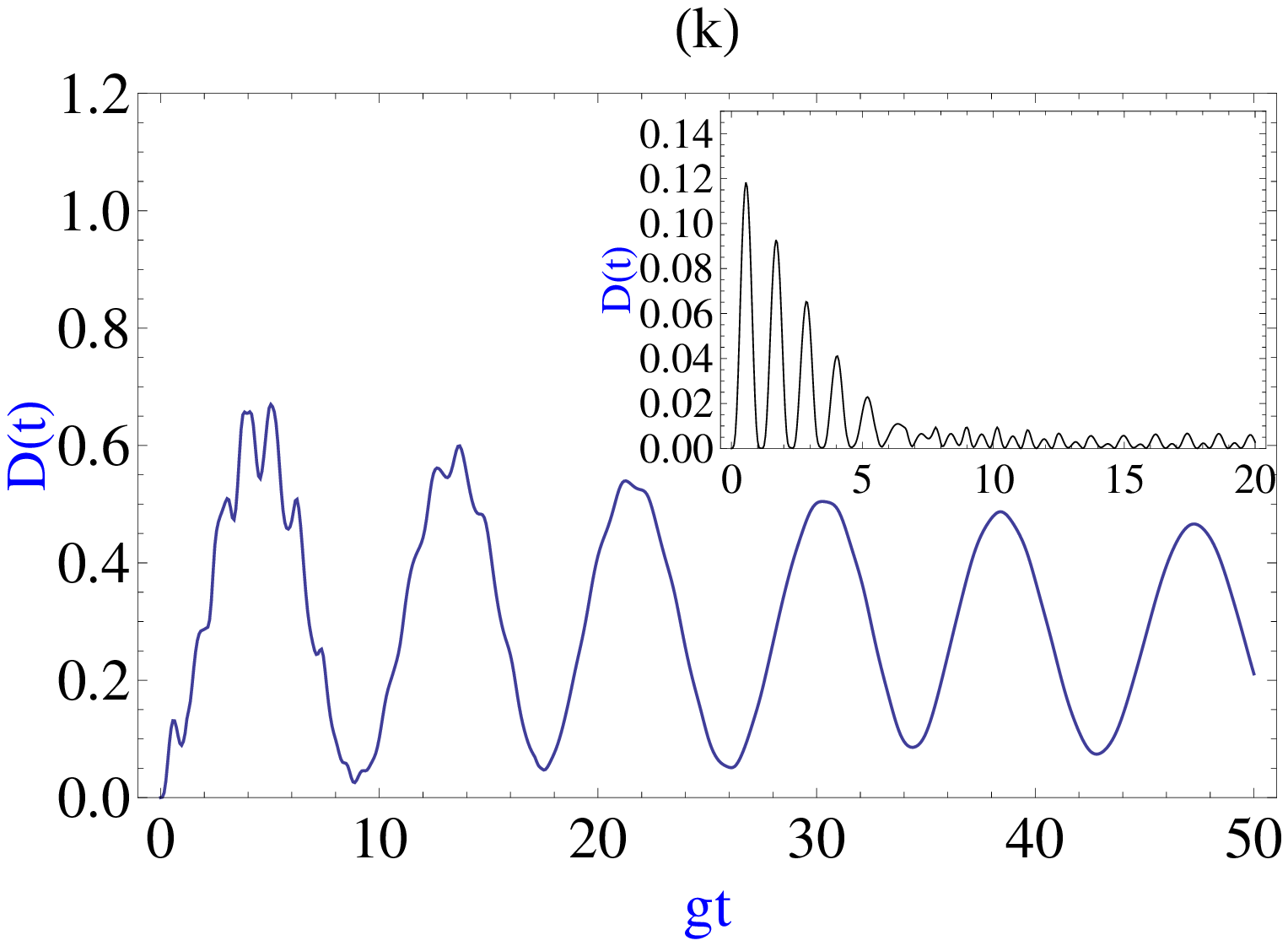}

\includegraphics[width=5.1cm]{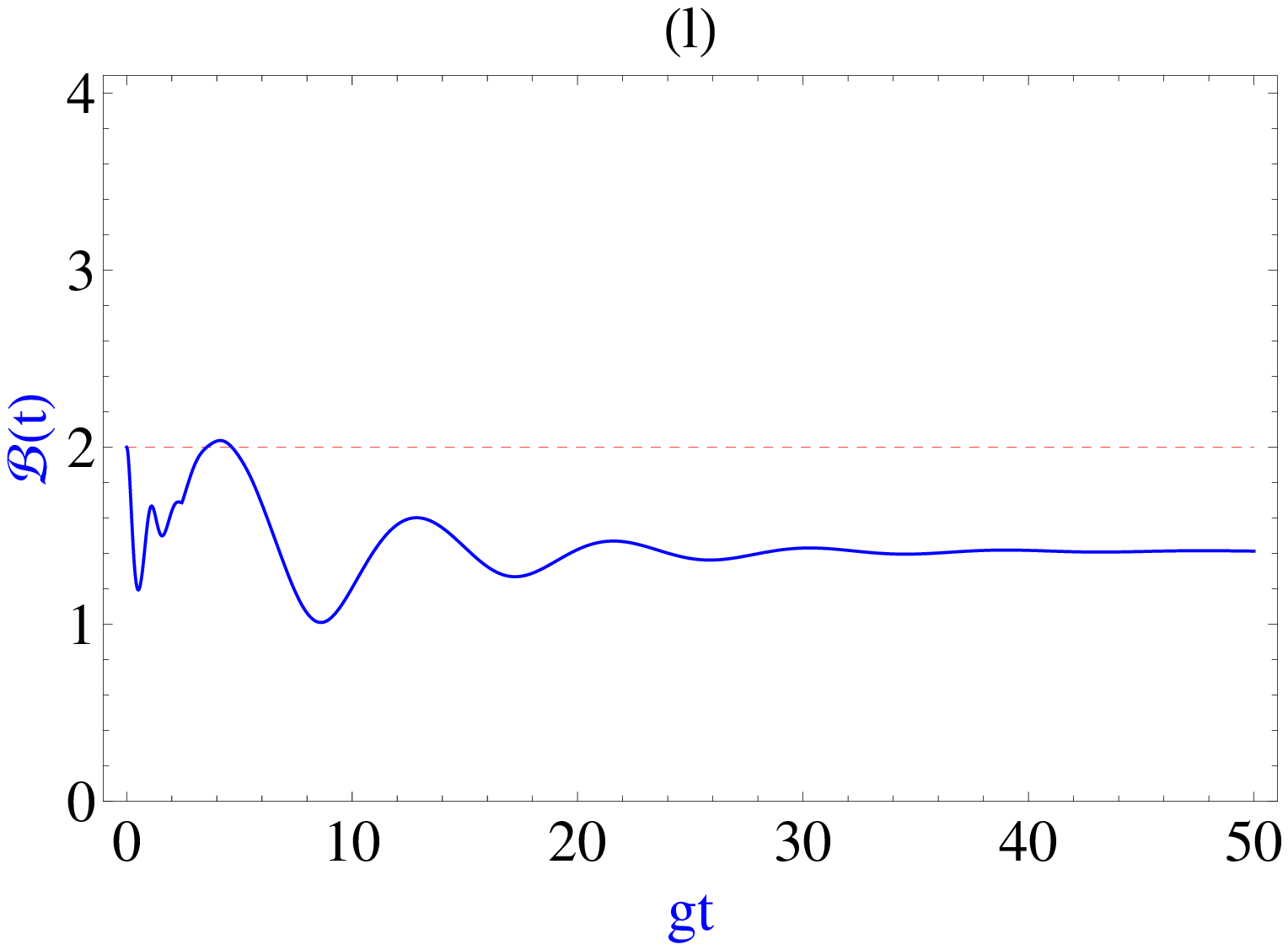}
\includegraphics[width=5.1cm]{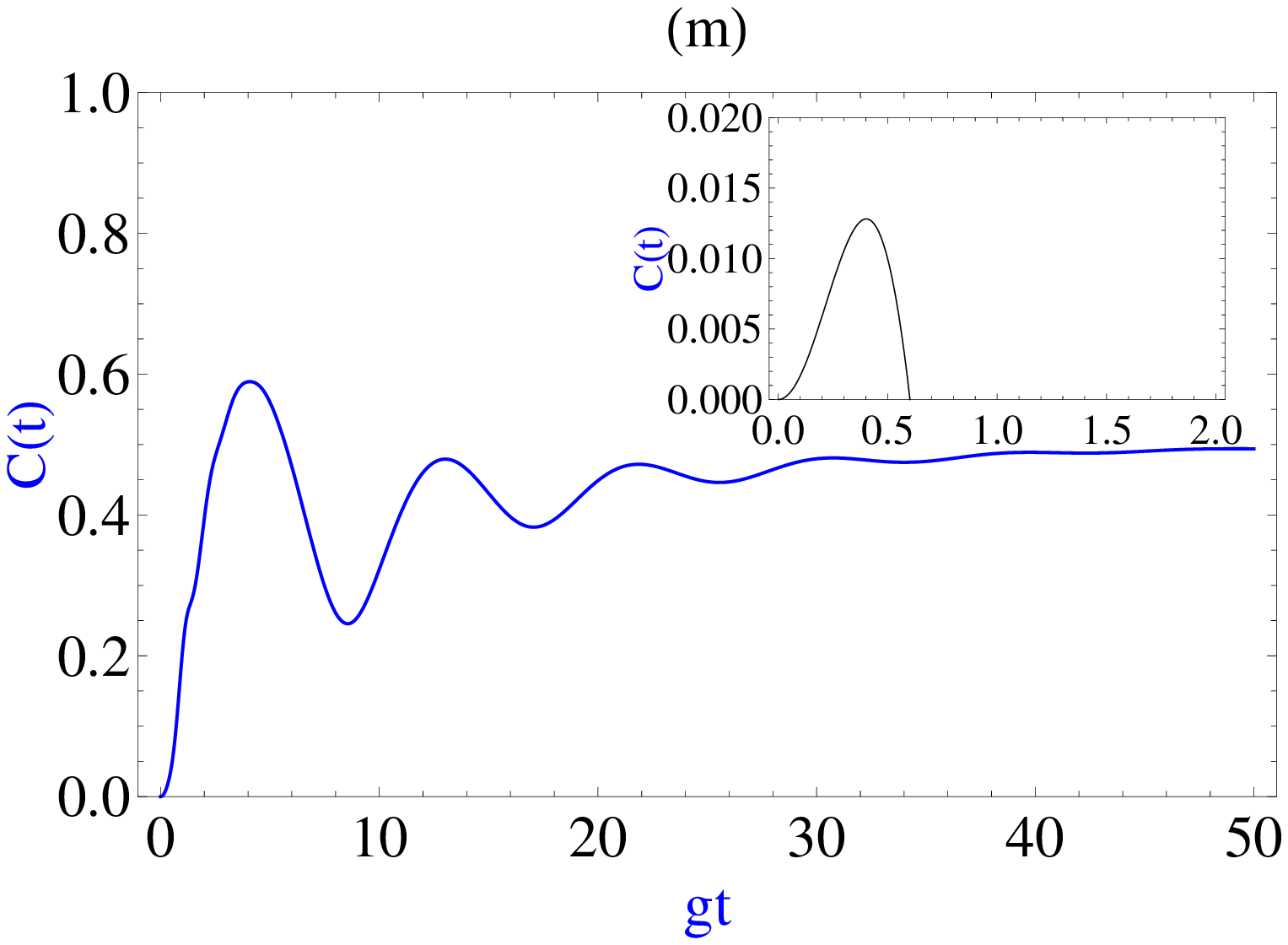}
\includegraphics[width=5.1cm]{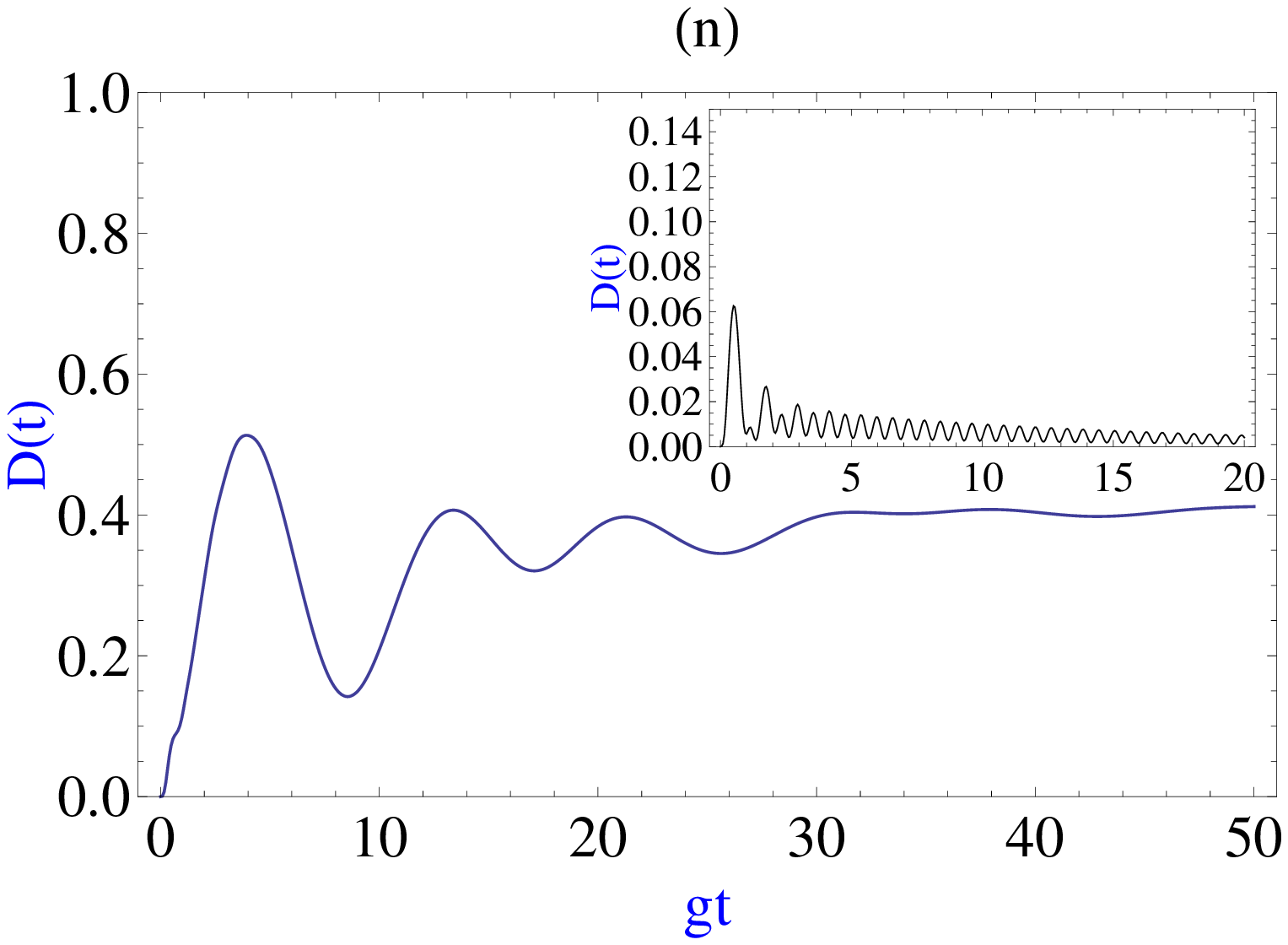}

\includegraphics[width=5.1cm]{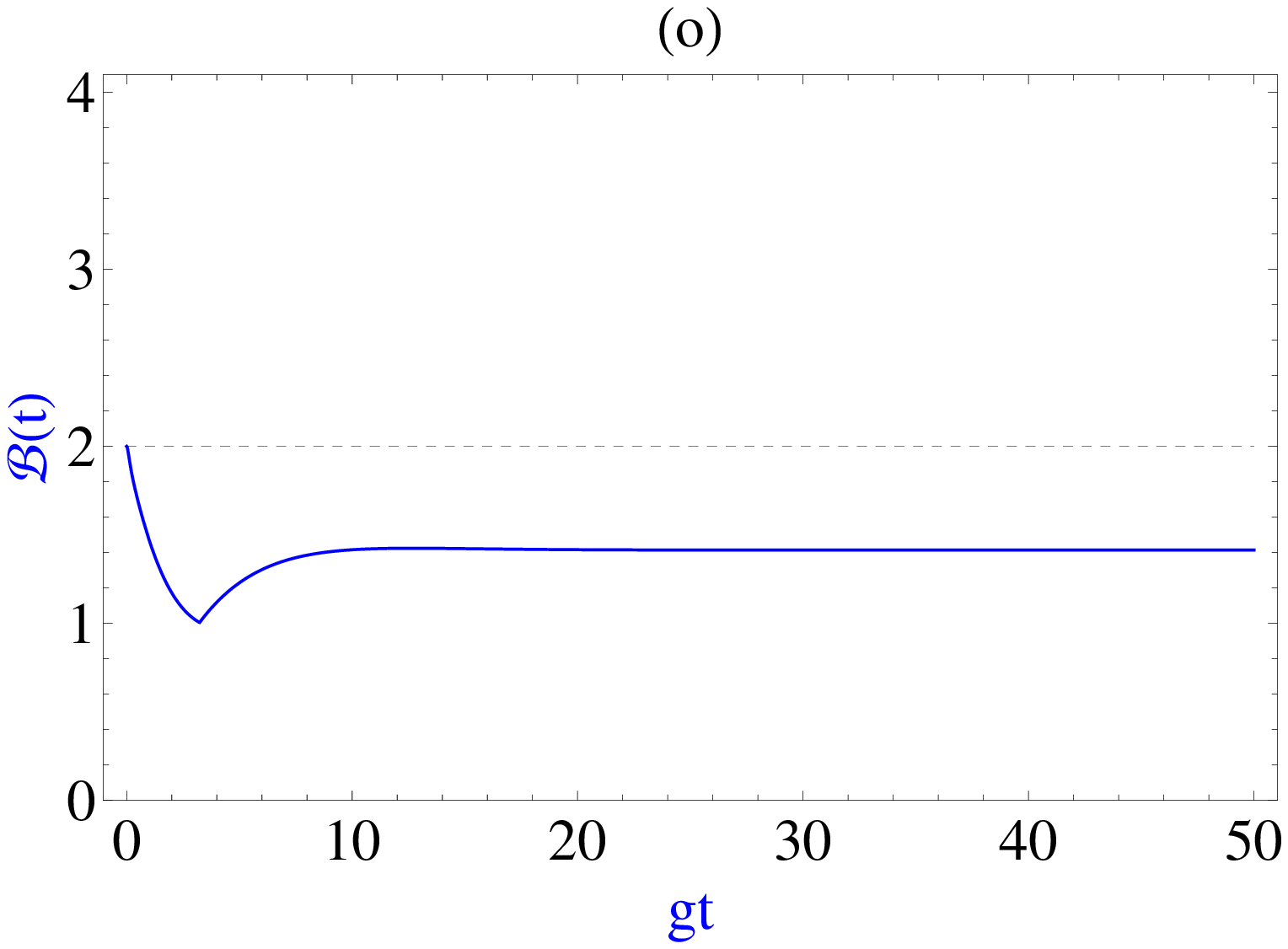}
\includegraphics[width=5.1cm]{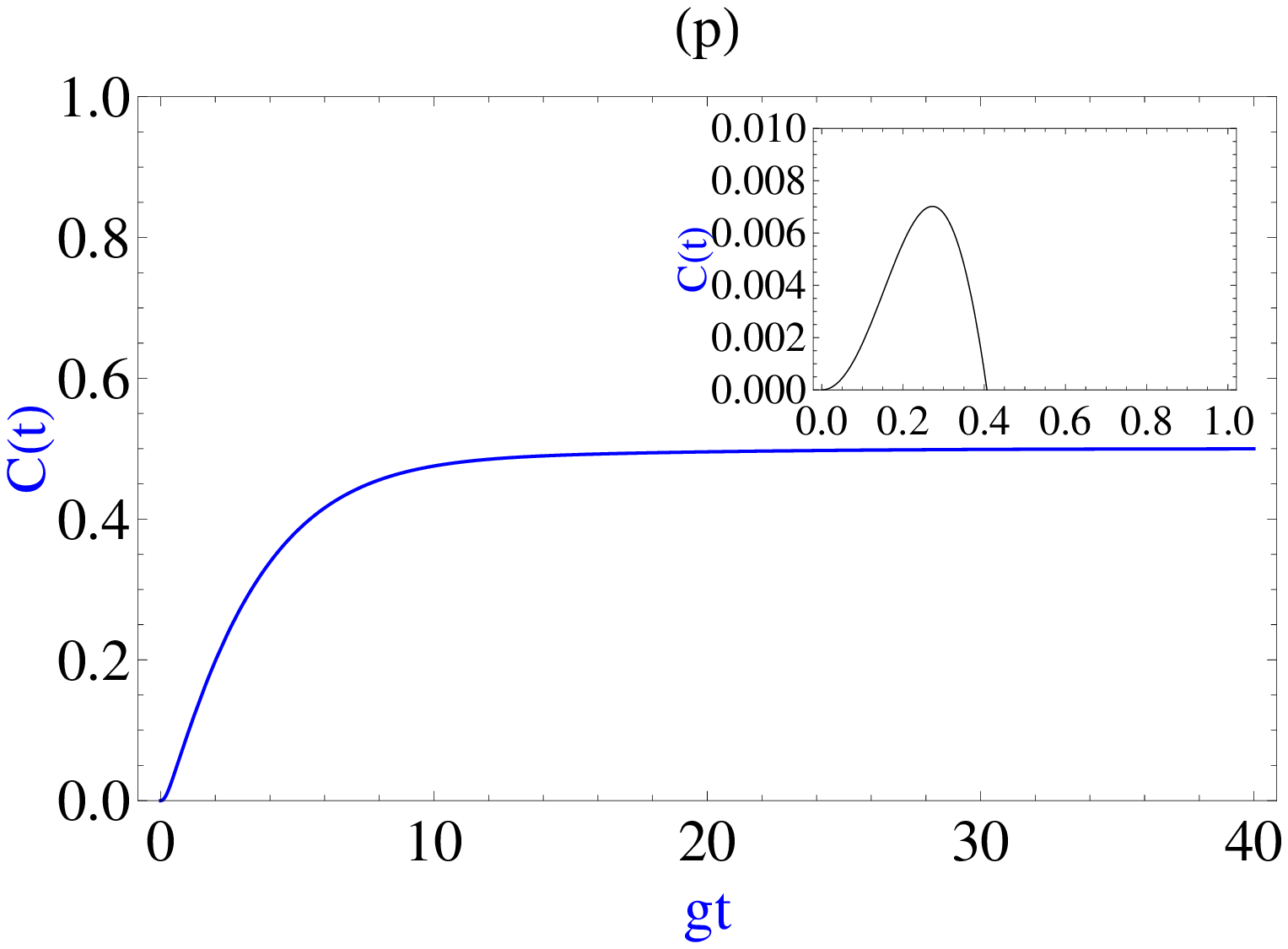}
\includegraphics[width=5.1cm]{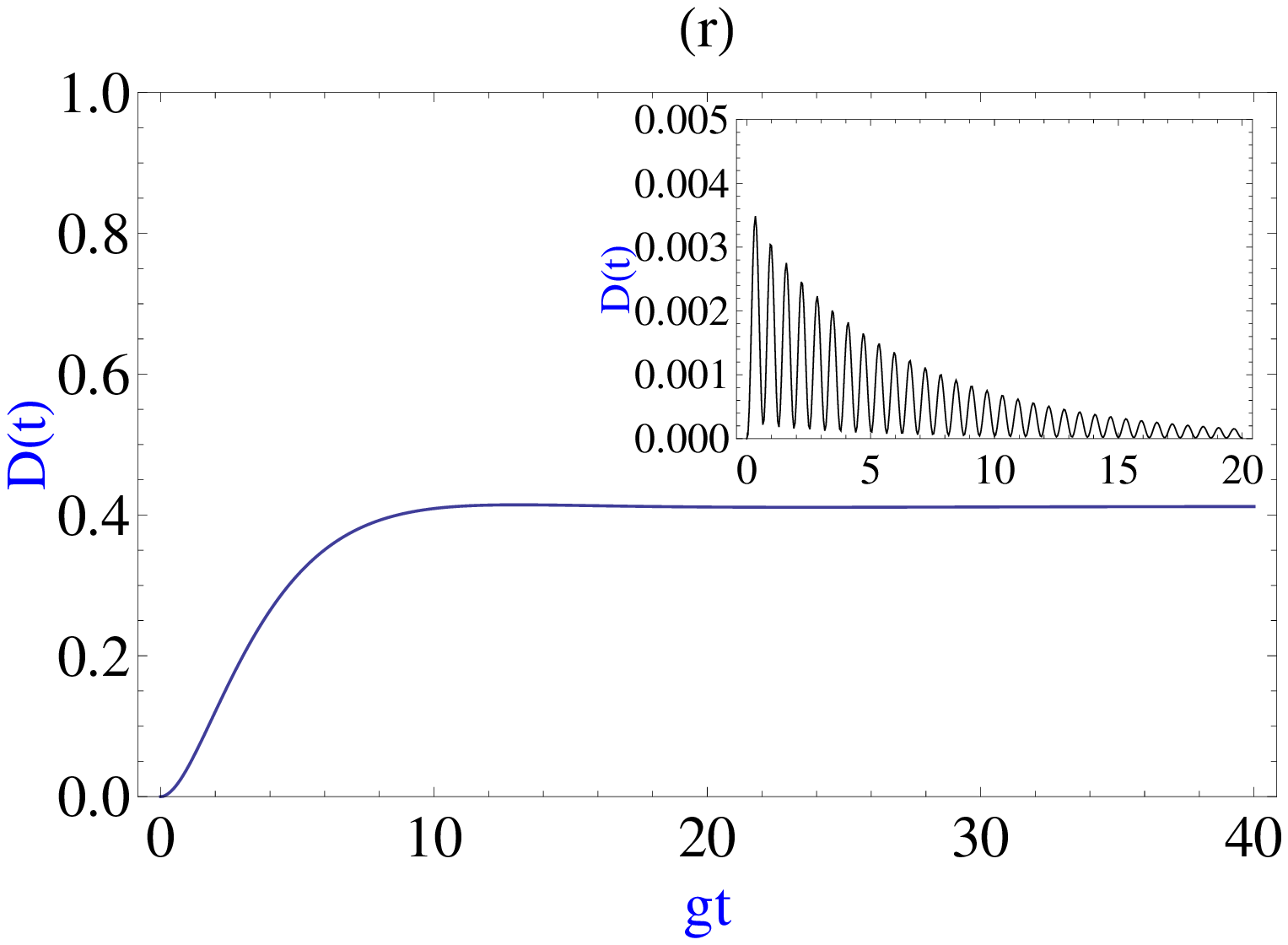}
\caption{\label{fig.3} CHSH inequality~(left plots), concurrence~(middle plots) and quantum discord~(right plots) versus $gt$ for $\rho(0)=\left|e_A,g_B,1\right\rangle\left\langle e_A,g_B,1\right|$ initial state and $\kappa=0$~((a),~(b),~(c)), $\kappa=0.02g$~((d),~(e),~(f)), $\kappa=0.2g$~((g),~(h),~(k)), $\kappa=2g$~((l),~(m),~(n)) and $\kappa=20g$~((o),~(p),~(r)). Here the main subfigures are for identical atoms with $\delta_1=\delta_2=5g$, while the insets are for unidentical atoms with $\delta_1=-\delta_2=5g$ under the same decay rate parameters. Note that for unidentical atoms $\mathcal{B}(t)$ is not violated, so they are not reported here.}
\end{figure}

Now we consider another initial separable state of the form $\rho(0)=\left|e_A,g_B\right\rangle\left\langle e_A,g_B\right|\otimes\left|1\right\rangle\left\langle 1\right|$ which has one excitation in the field and the other excitation in one of the atoms to see the effect of initial partition of the excitation on the dynamics of the quantum correlations. The time dependent of CHSH-inequality, concurrence and quantum discord are displayed in the first, second and third columns of Fig.~3, respectively. In these figures, insets show the dynamics for unidentical atoms~($\delta_1=-\delta_2=5g$) while main subfigures are for the identical atoms~($\delta_1=\delta_2=5g$). Inspection of subfigures demonstrates the pronounced effect of qubits being identical or unidentical. For identical atoms, the atom-field interaction induces high values of $C(t)$ and $D(t)$ and Bell inequality is also violated at the high values of entanglement for a high-quality cavity~(Fig.~3(a),~3(b) and~3(c)). In the case of unidentical atoms, $C(t)$ and $D(t)$ are also created, but their maximum are small; 0.018 and 0.13, respectively, and $\mathcal{B}(t)$ is not violated.  When unidentical atoms interact with a low-quality cavity, entanglement can cease to exist very quickly and QD shows a fast oscillatory decaying behaviour~(see the insets in Figs.~3(e)-3(r)) which is very similar to what we have found for $\rho(0)=\left|e_A,e_B,0\right\rangle\left\langle e_A,e_B,0\right|$ initial state. On the other hand, this situation is completely and significantly different in the case of identical atoms. Although CHSH-Bell inequality goes below 2 very quickly for $0<\kappa\leq 2.0g$ and no longer is violated for $\kappa>2.0g$, the dynamics of entanglement and quantum discord are very similar that both approach to  constant values in an oscillatory manner; the oscillations die down as the cavity decay rate is increased. In effect, cavity decay entangles initially separable state and the resulting entanglement as well as quantum discord are robust. In many studies, the dissipation of cavity fields were found to play a constructive role in the generation of quantum correlations as measured by concurrence and quantum discord~\cite{fpm,fcc,zcac,ykl,fmpp,mfzgp,hzsk}. This situation can be understood by looking at the steady-state of the dynamics given in Eq.~(\ref{mastereqn}). For $g_1=g_2=1$, $\kappa>0$ and $\delta_1=\delta_2=\delta$, the $\dot{\rho}=0$ equation has a solution independent of $\kappa$ and $\delta$ and the atom-atom reduced density matrix is given as: $\rho_{ss}=\frac{1}{4}(\left|e_A,g_B\right\rangle\left\langle e_A,g_B\right|+\left|g_A,e_B\right\rangle\left\langle g_A,e_B\right|)-\frac{1}{4}(\left|e_A,g_B\right\rangle\left\langle g_A,e_B\right|+\left|g_A,e_B\right\rangle\left\langle e_A,g_B\right|)+\frac{1}{2}\left|g_A,g_B\right\rangle\left\langle g_A,g_B\right|$. The concurrence for this state is 0.50 and quantum discord of it can be calculated from Eq.~(\ref{qd2}) to be 0.412 which are seen clearly from Figs.~3(p) and~3(r). Also, we want to emphasize that the dynamics of $\mathcal{B}$, $C$ and $D$ between identical atoms for the initial state $\left|\Psi(0)\right\rangle=\left|e_A,g_B,1\right\rangle$ is the same as that for $\left|\Psi(0)\right\rangle=\left|g_A,e_B,1\right\rangle$, since we assume symmetric coupling~(i.e., $g_1=g_2$) between the atoms and the single mode cavity field. 
\begin{figure}[!ht]\centering
\includegraphics[width=7.5cm]{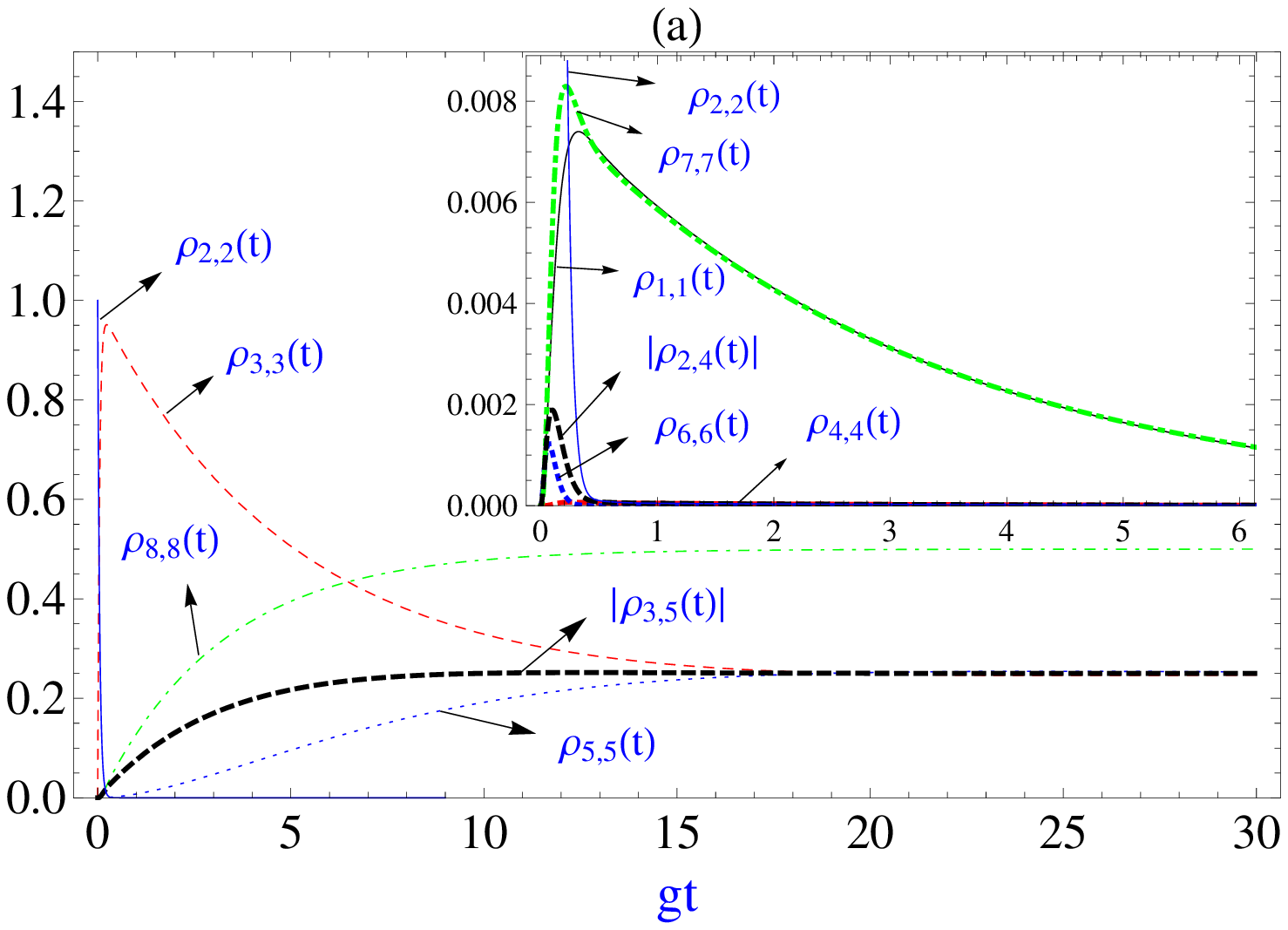}
\includegraphics[width=7.5cm]{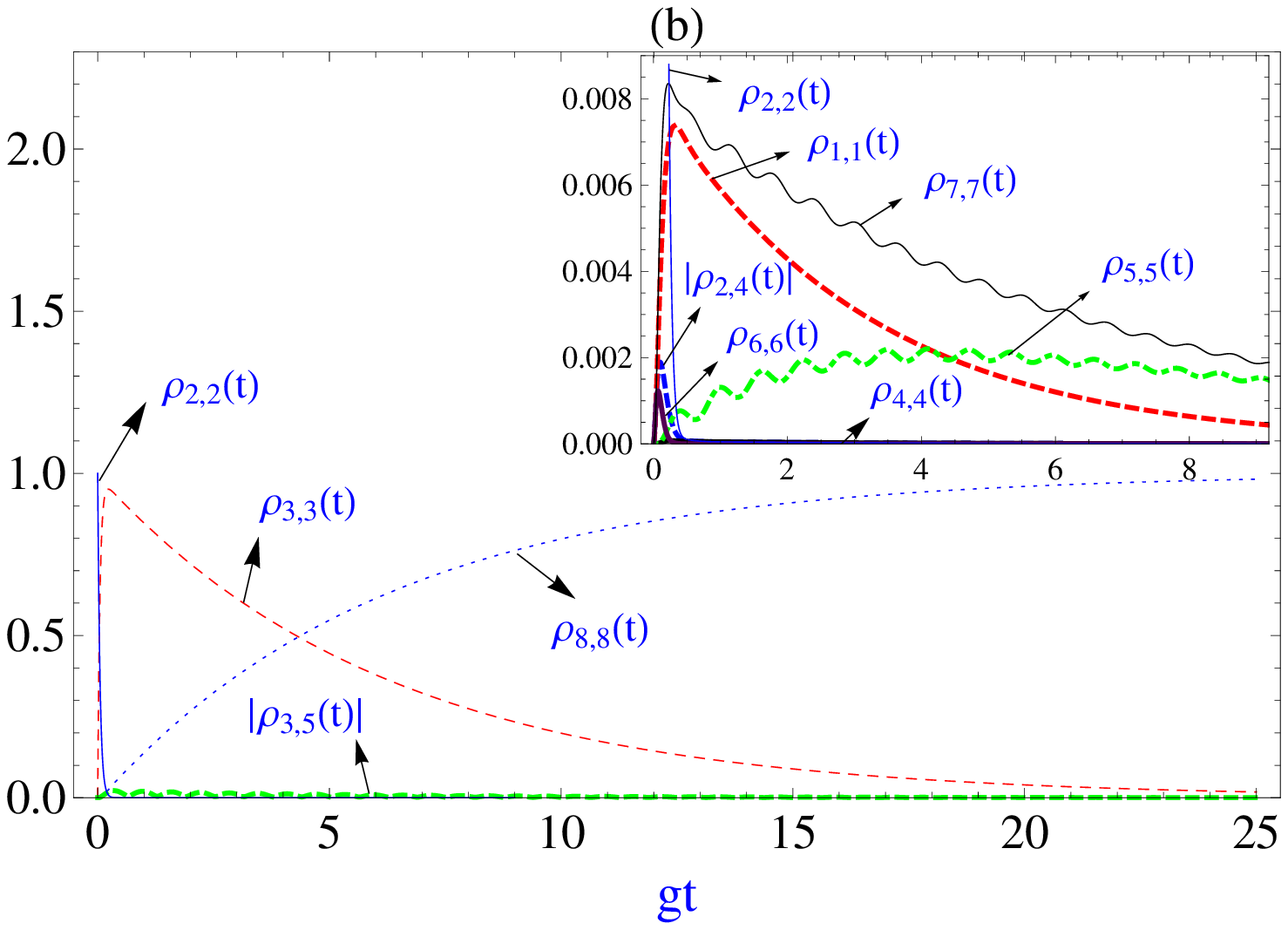}

\includegraphics[width=7.5cm]{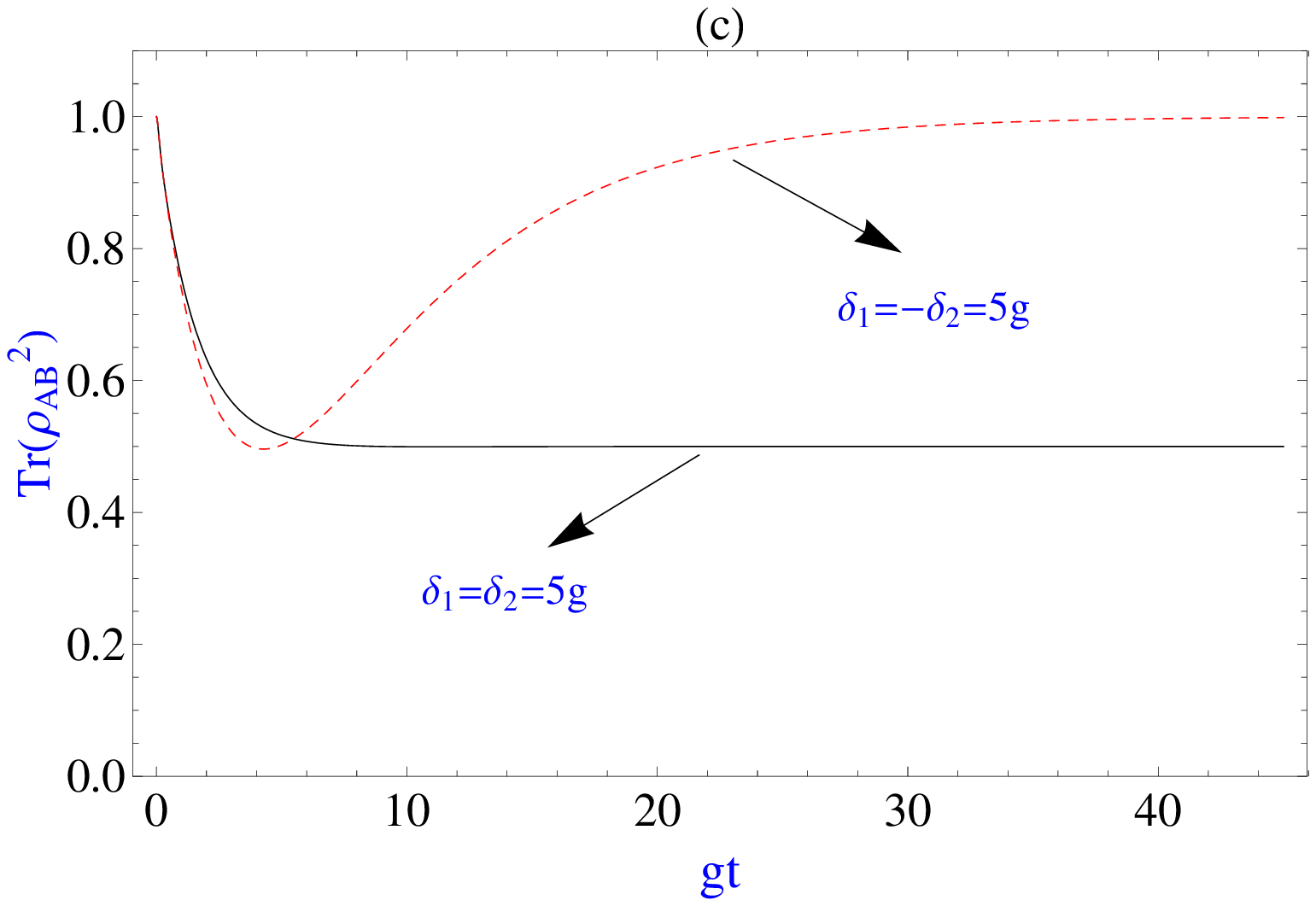}
\caption{\label{fig.4} The populations $\rho_{i,i}(t)$~$(i=1,2,...,8)$ and the absolute value of the coherences $\rho_{2,4}(t), \rho_{3,5}(t)$ given by Eq.~(\ref{solmaster}) versus $gt$ for the initial state $\rho(0)=\left|e_A,g_B,1\right\rangle\left\langle e_A,g_B,1\right|$ and $\kappa=20g$ for identical ($\delta_1=\delta_2=5g$)~(a) and unidentical ($\delta_1=-\delta_2=5g$)~(b) atoms. (c) shows the atomic purity~($Tr\rho_{AB}^2$) for the density matrix Eq.~(\ref{reducedden}) versus $gt$ for identical and unidentical atoms. Here the insets are for small time regions.}
\end{figure}

To further elucidate the asymptotic constant entanglement and quantum discord for the initial state $\rho(0)=\left|e_A,g_B,1\right\rangle\left\langle e_A,g_B,1\right|$, we display the time dependence of the density matrix elements of the atom-field system given by Eq.~(\ref{solmaster}) and the purity of the atomic subsystem  for the  density matrix Eq.~(\ref{reducedden}) for identical and unidentical detuning cases in Figs.~4(a)-4(c) for $\kappa=20g$. As can be seen from  Fig.~4(a), all populations, except $\rho_{3,3}, \rho_{5,5}$ and  $\rho_{8,8}$, become zero for a short time  while $\rho_{8,8}$ goes to $1/2$ and $\rho_{3,3}=\rho_{5,5}$ go to $1/4$ asymptotically. The key mechanism for the creation of the entanglement and QD in this case seems to be the coherence $\rho_{3,5}$ generated by the interaction of the atom with the dissipative cavity. The absolute value of $\rho_{3,5}$ also reaches to 1/4 asymptotically as can be seen from the Fig.~4(a). The evolved state is a trapping state which has the property that the probabilities of the different decay channels interfere destructively and the state is robust against the cavity loss mechanism~\cite{trapping,trapping2}. Generation or protecting entanglement by population trapping has been discussed by a number of groups before~\cite{fmpp,mfzgp,trapping,trapping2,trapping3}. The magnitude of $\kappa$ has no effect on the long time limit of $\rho_{3,3}, \rho_{5,5}$ and $\rho_{8,8}$, it only affects how long other population components can be nonzero. The dynamics of atomic purity~(which is equal to 1 for a pure state and 1/4 for a maximally mixed state in four dimensional Hilbert space)  displayed in Fig.~4(c) as the solid line for $\delta_1=\delta_2=5g$ indicates that the state approaches to a partially mixed state which  has high entanglement and quantum discord as indicated before in Figs.~3(p) and~3(r). For unidentical atoms, the short time dynamics is similar to the identical detuning case that all states are populated for a short time, but the asymptotic state is  now $\rho(t\rightarrow \infty )=\left|g_A,g_B,0\right\rangle\left\langle g_A,g_B,0\right|$ as can be seen from Fig.~4(b). The atomic purity (dotted line in Fig.~4(c)) indicates that the state of the atoms firstly goes to a mixed state in a short time and then approaches to the pure state $\rho=\left|g_A,g_B\right\rangle\left\langle g_A,g_B\right|$ which is the reason of the loss of correlations between unidentical atoms in the asymptotic limit.

\subsection{Effects of Cavity Decay-Maximally Entangled Initial State}\label{cavity2}
In this subsection, we will investigate the effect of initial correlations in the atom-atom part of the state on the dynamics of quantum correlations. 
\begin{figure}[!ht]\centering
\includegraphics[width=7.5cm]{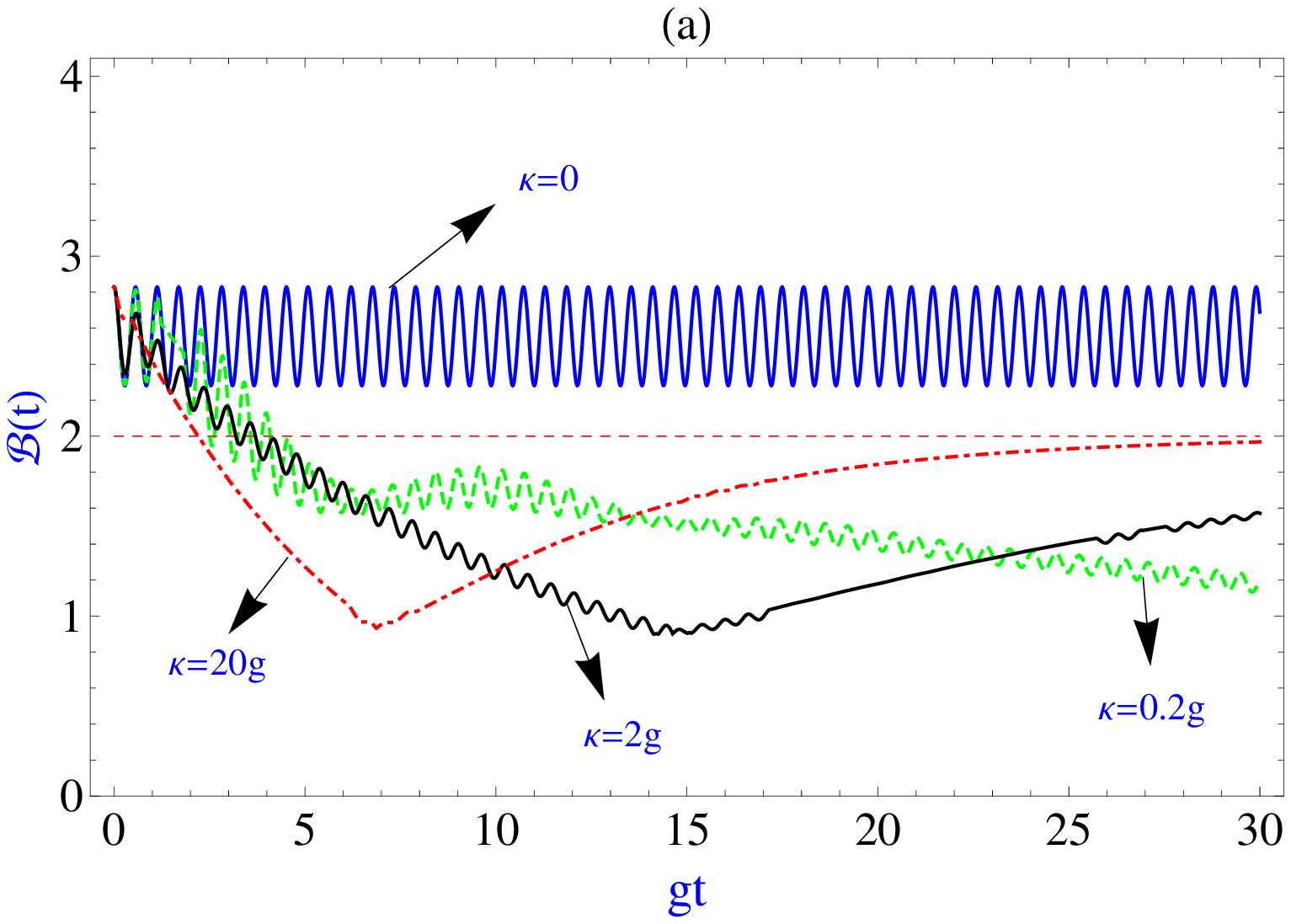}
\includegraphics[width=7.5cm]{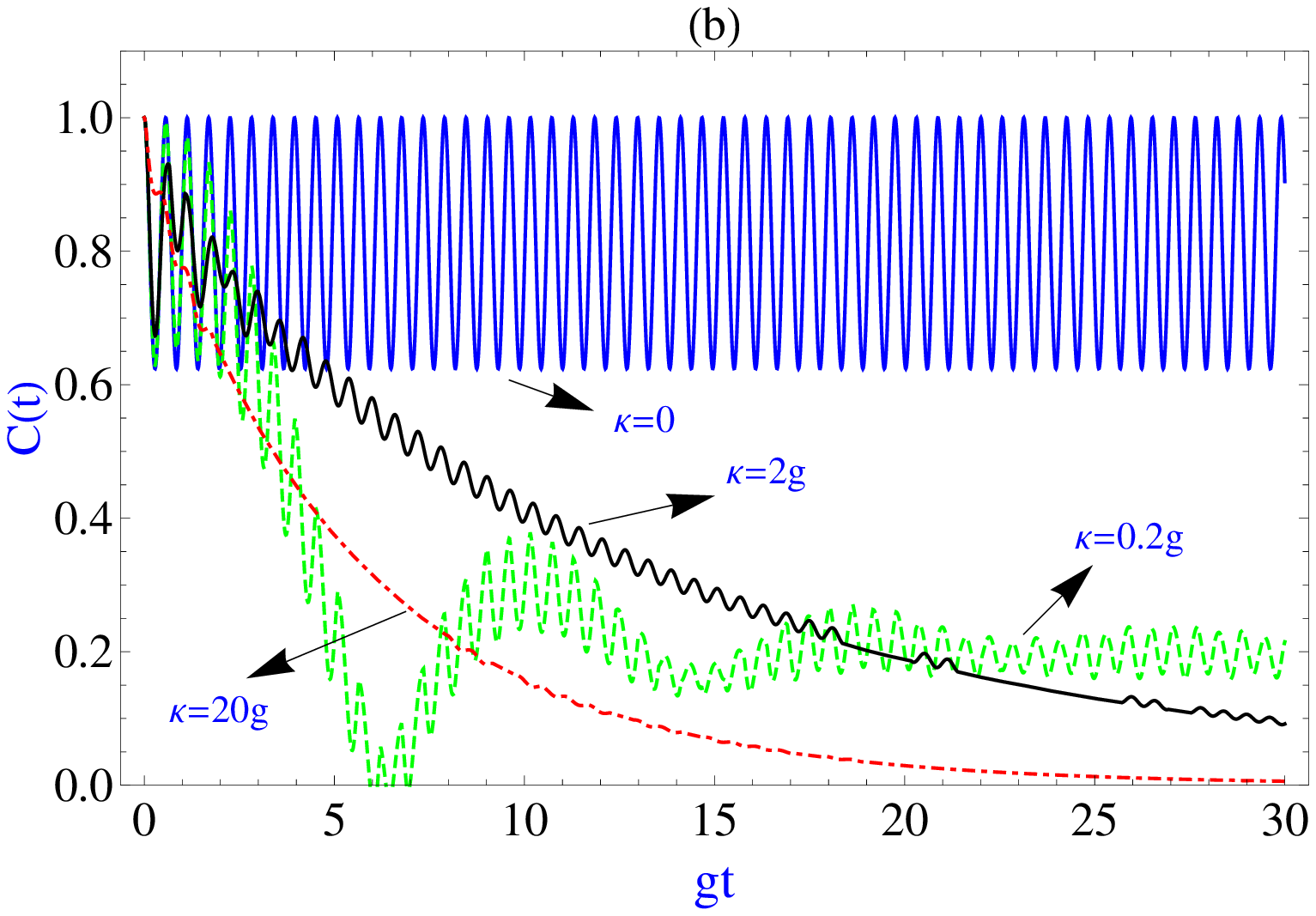}

\includegraphics[width=7.5cm]{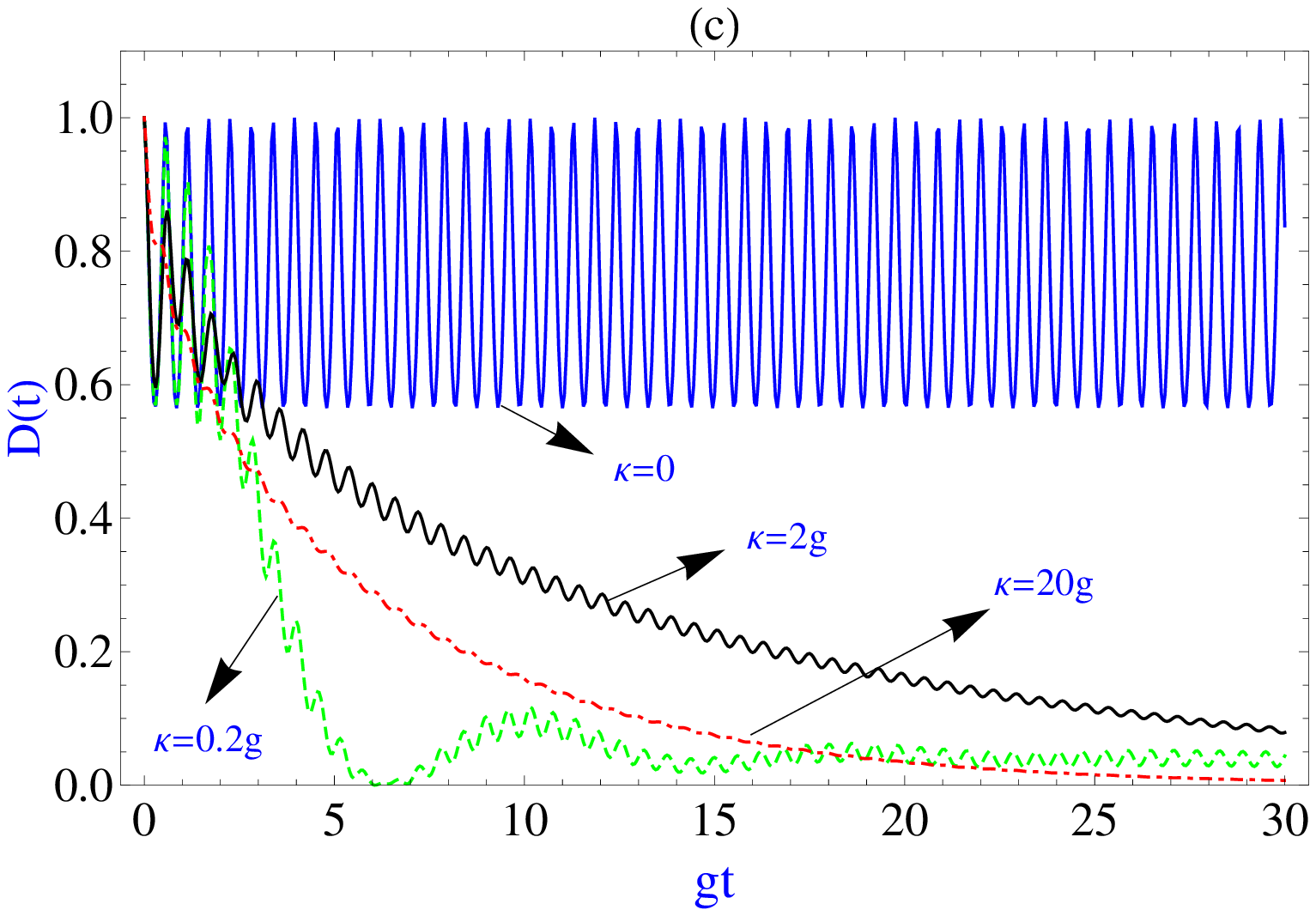}
\caption{\label{fig.5} CHSH inequality~(a), concurrence~(b) and quantum discord~(c) versus $gt$ for unidentical atoms, $\delta_1=-\delta_2=5g$ and $\rho(0)=\left|\Psi(0)\right\rangle\left\langle\Psi(0)\right|$ where $\left|\Psi(0)\right\rangle=\frac{1}{\sqrt{2}}(\left|g_A,e_B,1\right\rangle+\left|e_A,g_B,1\right\rangle)$. Here all subfigures include plots for $\kappa=0$, $\kappa=0.2g$, $\kappa=2g$ and  $\kappa=20g$.}
\end{figure}

We take $\left|\Psi(0)\right\rangle=\frac{1}{\sqrt{2}}(\left|g_A,e_B\right\rangle+\left|e_A,g_B\right\rangle)\otimes\left|1\right\rangle$ as the initial state which is the maximally entangled Bell state for the atomic system. QD and Bell inequality violation are also maximal for this initial state. We display $\mathcal{B}$, $C$ and $D$ as a function of $gt$ for several cavity decay rates for unidentical~(Fig.~5) and identical~(Fig.~6) atoms. At $\kappa=0$, all correlations show oscillatory behaviour; for the unidentical atoms $\mathcal{B}(t)$, $C(t)$ and $D(t)$ oscillate nearly in the ranges [$2\sqrt{2}$, 2.2], [1.0, 0.63] and [1.0, 0.58], respectively~(Fig.~5). On the other hand, $\mathcal{B}(t)$ and $C(t)$ for $\kappa=0$ have periodic sudden deaths and births for the identical detunings of the atoms~(Figs.~6(a) and~6(b)), while $D(t)$ oscillates nearly between 0.3 and 1~(Fig.~6(c)). For small values of $\kappa$~$(\kappa=0.2g)$, the effect of cavity decay on the dynamics of correlations can be described as underdamped oscillations for both identical and unidentical detuning cases. As $\kappa$ is increased to $2g$, the oscillation becomes overdamped and finally at large $\kappa$'s~($\kappa=20g$) damping becomes critical, i.e. $C(t)$ and $D(t)$ show exponential decay for the unidentical atom cases. For the identical atoms at $\kappa=2g$ and $\kappa=20g$ concurrence has sudden death, while QD decays exponentially. One should note that Bell nonlocality is the most fragile quantum correlation considered here; it suffers sudden death in a short time as long as $\kappa>0$.
\begin{figure}[!ht]\centering
\includegraphics[width=5.1cm]{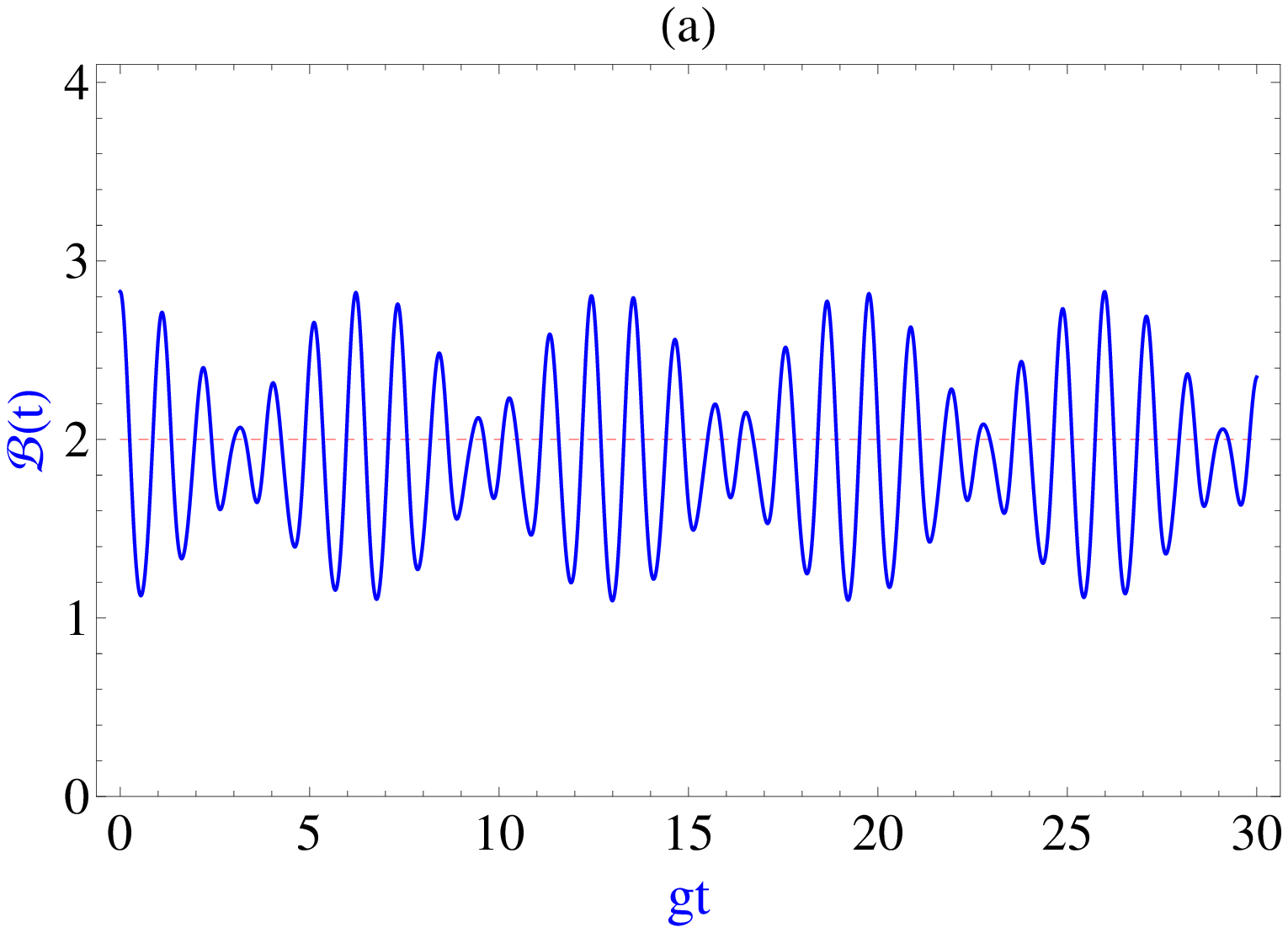}
\includegraphics[width=5.1cm]{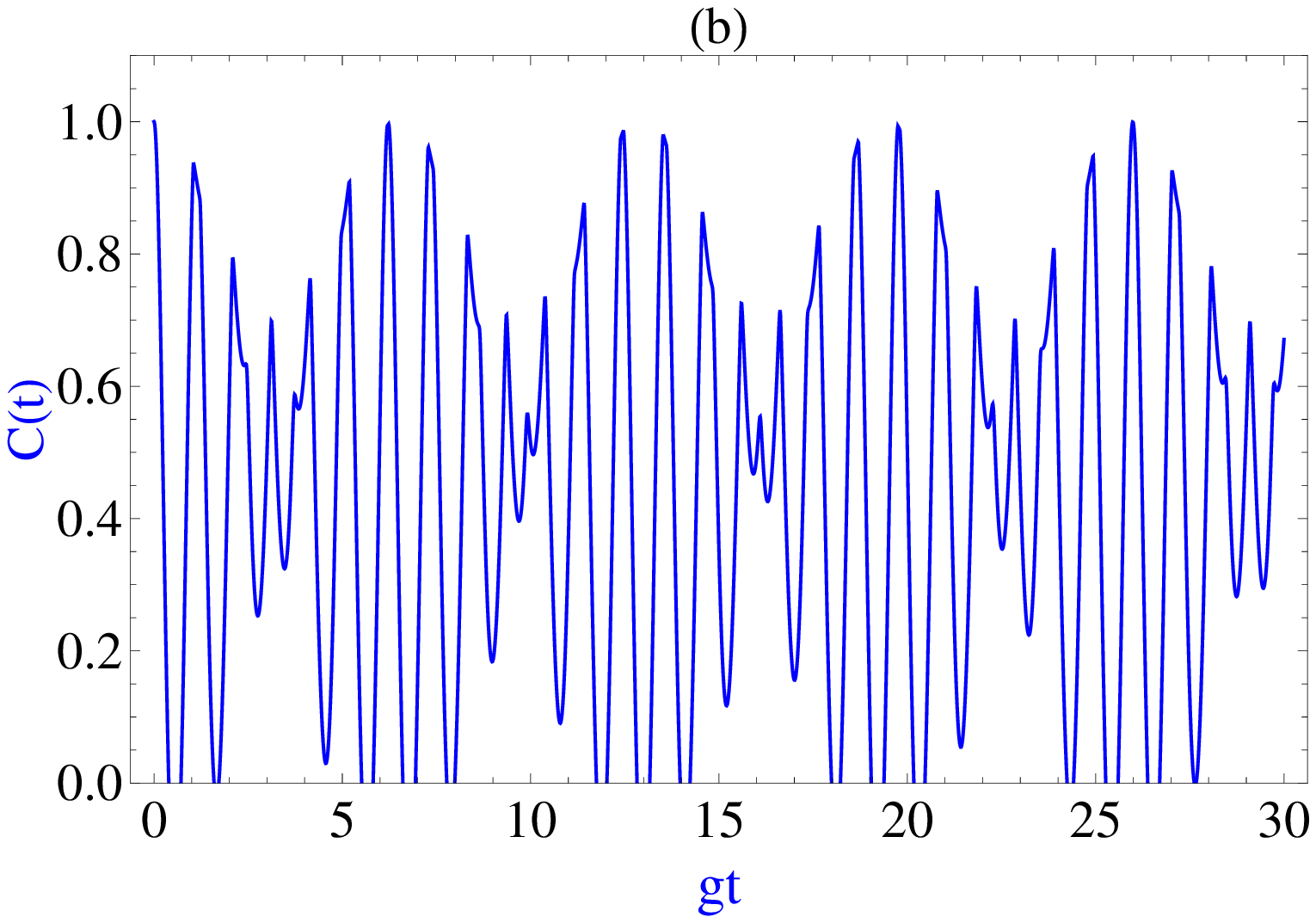}
\includegraphics[width=5.1cm]{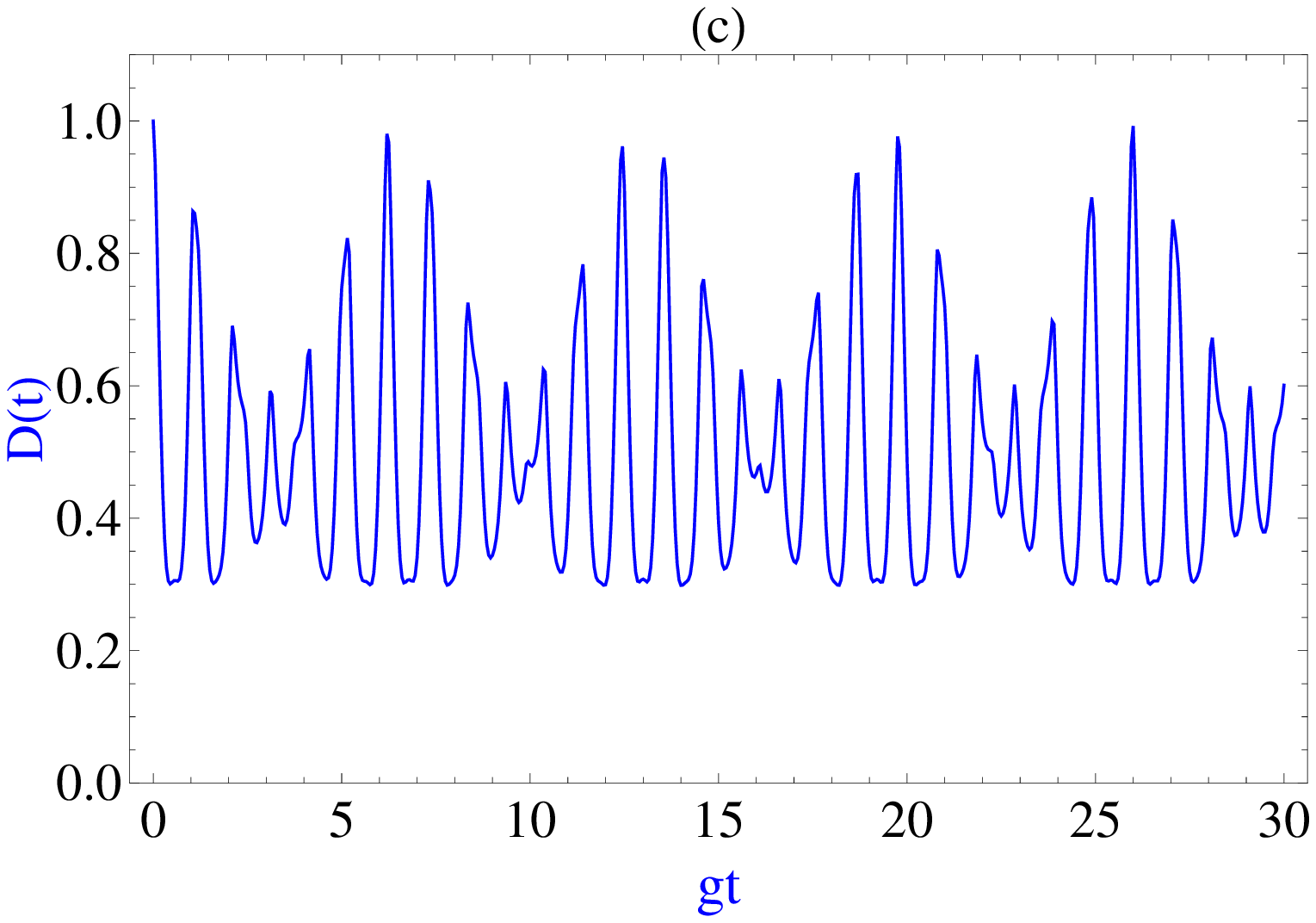}

\includegraphics[width=5.1cm]{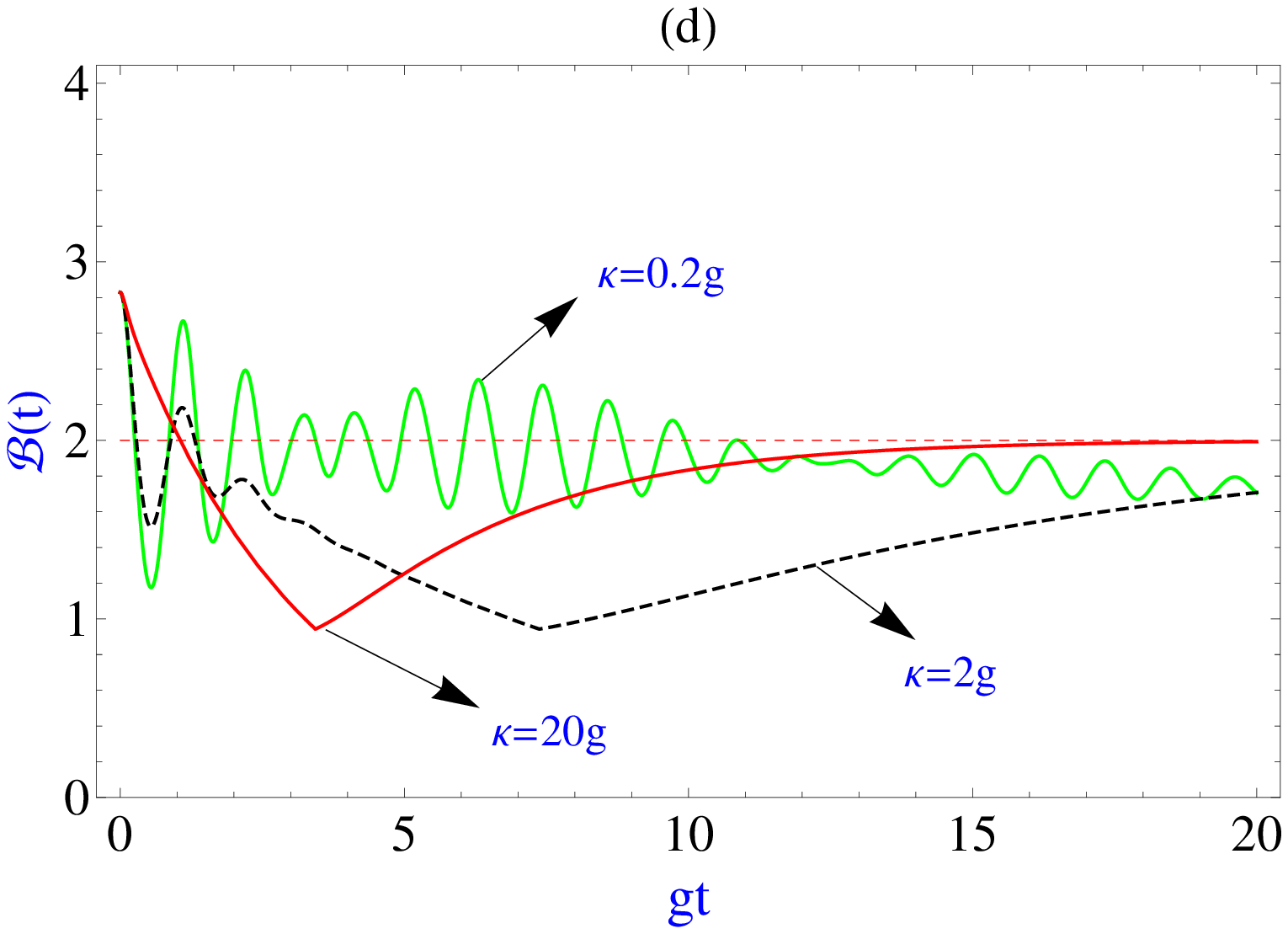}
\includegraphics[width=5.1cm]{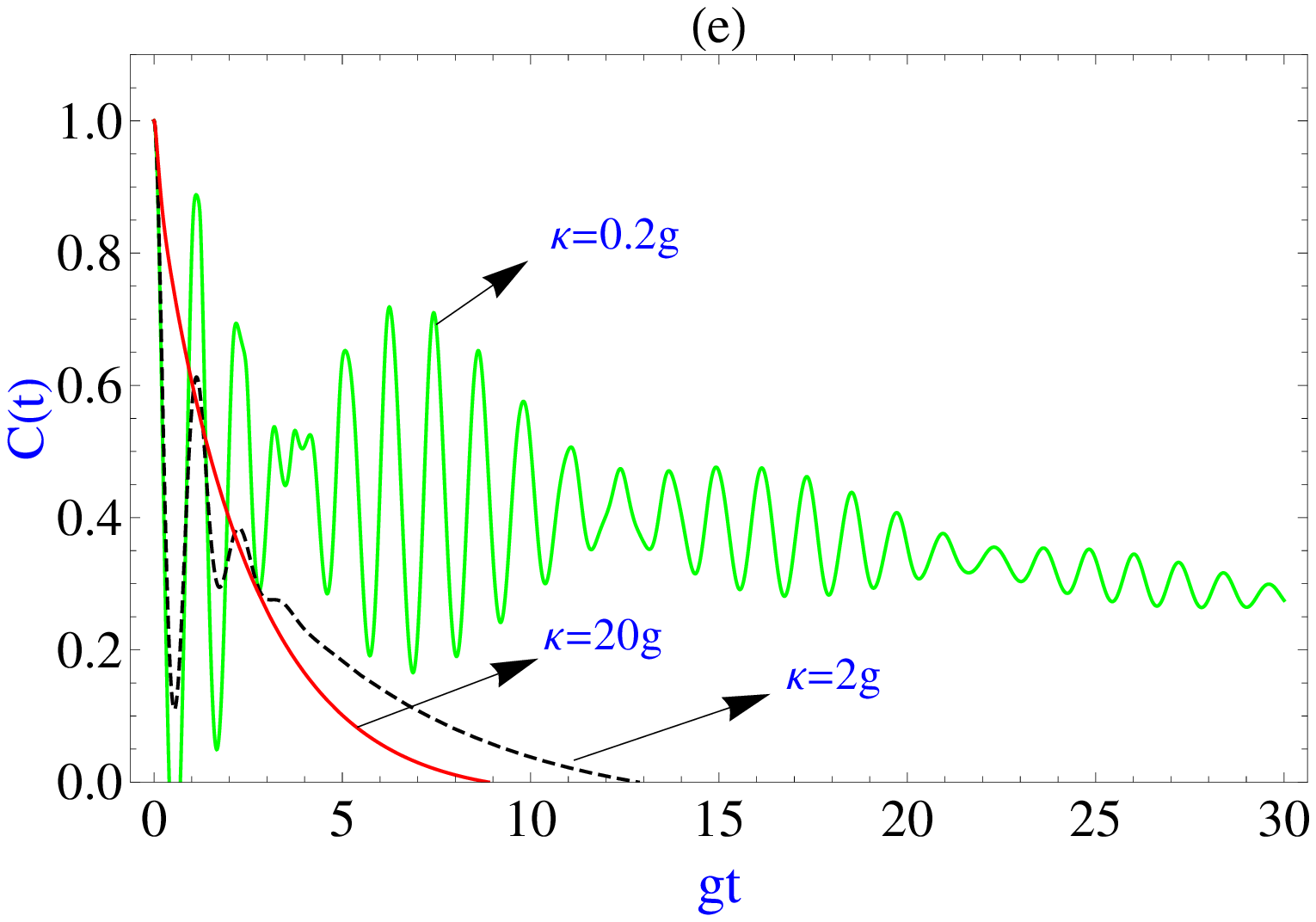}
\includegraphics[width=5.1cm]{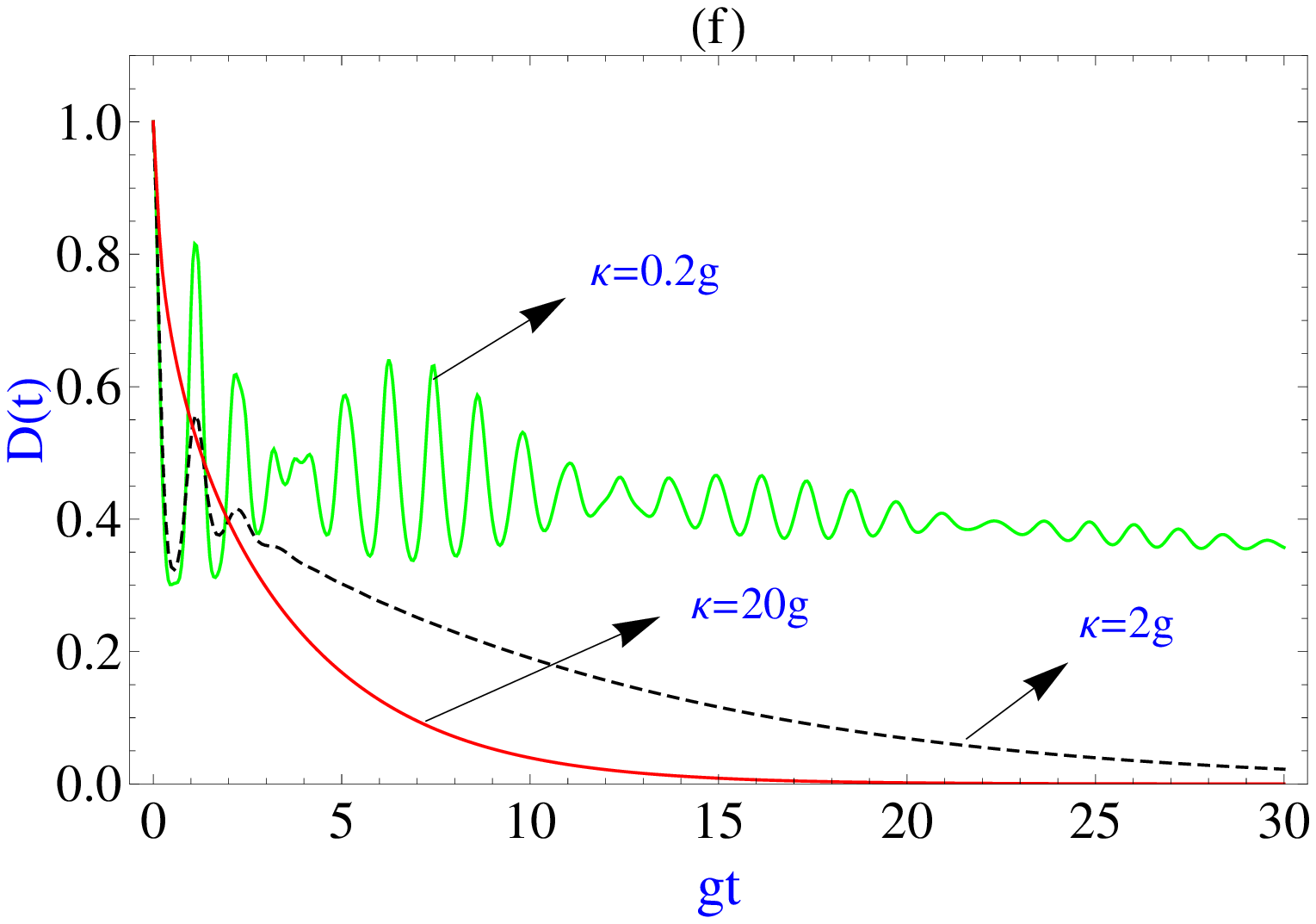}
\caption{\label{fig.6} CHSH inequality~(a) and~(d), concurrence~(b) and~(e) and quantum discord~(c) and~(f) versus $gt$ for identical atoms, $\delta_1=\delta_2=5g$ and $\rho(0)=\left|\Psi(0)\right\rangle\left\langle\Psi(0)\right|$ where $\left|\Psi(0)\right\rangle=\frac{1}{\sqrt{2}}(\left|g_A,e_B,1\right\rangle+\left|e_A,g_B,1\right\rangle)$. Here (a), (b) and (c) are for $\kappa=0$, while (d), (e) and (f) include plots for $\kappa=0.2g$, $\kappa=2g$ and  $\kappa=20g$.}
\end{figure}

\section{Conclusion}\label{conclusion}  
We have investigated the dynamics of quantum correlations as quantified by entanglement, Bell nonlocality and quantum discord for two qubits embedded in a common leaky cavity for three different initial states with double excitations. The qubits were considered as identical or unidentical based on the detuning of their transition frequency with respect to the frequency of the field mode. The main findings of the study can be summarized as:
Qubits being identical or not have profound influence on the dynamics of the considered correlations; starting from the same initial state, the correlations show very different behaviour depending on whether the detunings of the individual qubits satisfy $\delta_1=\delta_2=\delta$ or  $\delta_1=-\delta_2=\delta$. For unidentical qubits initially in their excited states and interacting with a high or low quality cavity, quantum discord is found to be induced for all considered cavity decay rates and it decays oscillatorily for a leaky cavity although the qubits remain unentangled for all times. For the same initial state and the leaky cavity, QD is found to increase in the long time for identical qubits, while entanglement is either "never created" or decays to zero in a short time.

One of the most interesting  finding is the constant high entanglement and quantum discord state which is obtained as the long-time limit of the dynamics of $\rho(0)=\left|e_A,g_B\right\rangle\left\langle e_A,g_B\right|\otimes\left|1\right\rangle\left\langle 1\right|$  initial state for the identical qubit case. Here the final state is independent of the cavity decay rate~(if $\kappa>0$) and the magnitude of the detuning and has high concurrence~($C=0.5$) and quantum discord~($D=0.412$). From an analysis of the time dependence of the density matrix elements it is seen that the considered initial state leads to a trapping state which happen to have high degree of quantum correlations. On the other hand, for the same initial state, the correlations between the unidentical atoms are found to be destroyed quickly even for a small cavity decay rate.

For the atomic Bell-type initial state we have found ESD for the identical qubits interacting with a low quality cavity, while concurrence decays exponentially for the unidentical qubits. Quantum discord is found to have asymptotical decay behaviour independent of the type of atoms for this initial state and leaky cavity. On the other hand, CHSH-Bell inequality violations survive only for a finite time for $\kappa>0$ for both identical and unidentical qubits. It is interesting to note that steady-state nonzero correlation state of  $\left|\Psi(0)\right\rangle=\left|e_A,g_B\right\rangle\otimes\left|1\right\rangle$~(or  $\left|\Psi(0)\right\rangle=\left|g_A,e_B,1\right\rangle$) is not observed for the   $\left|\Psi(0)\right\rangle=\frac{1}{\sqrt{2}}(\left|g_A,e_B\right\rangle+\left|e_A,g_B\right\rangle)\otimes\left|1\right\rangle$. This shows that the stationary states depend strongly on the initial conditions~\cite{mfzgp}. 

We have also demonstrated that Bell nonlocality as manifested by CHSH inequality is the most fragile correlation measure under dissipative environment and is violated only at the high values of entanglement.

\end{document}